\documentclass[a4paper, twocolumn, 11pt]{quantumarticle}
\pdfoutput = 1
\usepackage[english]{babel}
\usepackage[a4paper,top=2cm,bottom=2cm,left=2.5cm,right=2.5cm,marginparwidth=1.75cm]{geometry}

\usepackage{setspace}

\usepackage{amsmath, amssymb}
\usepackage{mathtools}
\usepackage{graphicx}
\usepackage{overpic}
\usepackage[shortlabels]{enumitem}
\usepackage[colorlinks=true, allcolors=blue]{hyperref}
\usepackage{chemformula} 
\usepackage{pythonhighlight}  
\usepackage{comment} 
\usepackage{tablefootnote} 
\usepackage{multirow}
\usepackage{ulem}

\usepackage{physics}
\usepackage{subcaption}
\usepackage{tikz}
\usetikzlibrary{arrows.meta, positioning, shapes.geometric, shadows}
\usepackage{standalone}
\usepackage{amsthm}

\usepackage[numbers,sort&compress]{natbib}

\usepackage{cleveref}
\crefname{theorem}{Theorem}{Theorems}
\crefname{section}{Section}{Sections}
\crefname{equations}{Eq.}{Eqs.}

\usepackage{listings}
\usepackage{xcolor}

\definecolor{codegreen}{rgb}{0,0.6,0}
\definecolor{codegray}{rgb}{0.5,0.5,0.5}
\definecolor{codepurple}{rgb}{0.58,0,0.82}
\definecolor{backcolour}{rgb}{0.95,0.95,0.92}

\lstdefinestyle{mystyle}{
    backgroundcolor=\color{backcolour},   
    commentstyle=\color{codegreen},
    keywordstyle=\color{magenta},
    numberstyle=\tiny\color{codegray},
    stringstyle=\color{codepurple},
    basicstyle=\ttfamily\footnotesize,
    breakatwhitespace=false,         
    breaklines=true,                 
    captionpos=b,                    
    keepspaces=true,                 
    numbers=left,                    
    numbersep=1pt,                  
    showspaces=false,                
    showstringspaces=false,
    showtabs=false,                  
    tabsize=2
}

\usepackage{mathtools}

\begin{document}

\title{Ans\"atz Expressivity and Optimization in Variational Quantum Simulations of Transverse-field Ising Model Across System Sizes }

\author{Ashutosh P. Tripathi}
\email{ashutosh.tripathi@tifr.res.in}
\affiliation{Department of Theoretical Physics, Tata Institute of Fundamental Research, \\Homi Bhabha Road, Mumbai 400005, India}

\author{Nilmani Mathur}
\email{nilmani@theory.tifr.res.in}
\affiliation{Department of Theoretical Physics, Tata Institute of Fundamental Research, \\Homi Bhabha Road, Mumbai 400005, India}

\author{Vikram Tripathi}
\email{v.tripathi@theory.tifr.res.in}
\affiliation{Department of Theoretical Physics, Tata Institute of Fundamental Research, \\Homi Bhabha Road, Mumbai 400005, India}



\begin{abstract}
We explore the application of the Variational Quantum Eigensolver (VQE) to investigate the ground state properties, particularly the entanglement entropy, of the Transverse Field Ising Model (TFIM) in one, two, and three dimensions, considering systems of up to 27 spins. By benchmarking VQE results against exact diagonalization and analyzing the entanglement properties across different system sizes
, we assess the algorithm's effectiveness in capturing 
faithful ground state. Using results of TFIM, we also investigate how VQE's expressivity and optimization influence the simulation of highly entangled quantum states.
We employ different ans\"atze: the hardware-efficient EfficientSU2 from Qiskit, the physics-inspired Hamiltonian Variational ans\"atz (HVA) and HVA with symmetry breaking, and benchmark their performance using energy variance, entanglement entropy, spin correlations, and magnetization. We further discuss the implications for scaling these methods to larger quantum systems.

\end{abstract}

\maketitle

\section{\label{sec:intro}Introduction}
The investigation of quantum many-body systems is a very important area of modern physics, comprising a wide range of phenomena in condensed matter, nuclear and high energy physics, and in general for strongly interacting quantum field theories. In practice, realistic systems are often intractable, and one relies on simplified models that capture essential features of the underlying physics. Over the past several decades, classical numerical methods such as exact diagonalization (ED), density matrix renormalization group (DMRG), quantum Monte Carlo (QMC), and tensor-network approaches have led to significant progress in understanding these systems. However, their applicability is fundamentally limited by the exponential growth of Hilbert space, the emergence of sign problems in fermionic and frustrated systems, and the difficulty of efficiently representing highly entangled states, particularly in higher spatial dimensions. 

Quantum computing offers a promising alternative approach by enabling the direct simulation of quantum systems on quantum hardware. In the current era of noisy intermediate-scale quantum (NISQ) devices, which now exceed 1000 qubits \cite{castelvecchi_IBM_2023, acharya_2024, Nation_2025}, there is considerable interest in developing algorithms that can operate within the constraints of limited coherence times, gate errors, and shallow circuit depths. Among these, the Variational Quantum Eigensolver (VQE) has emerged as a leading candidate for near-term quantum simulation \cite{Peruzzo_2014}. VQE is a hybrid classical-quantum algorithm in which a parameterized quantum circuit prepares a trial state, and a classical optimizer iteratively minimizes the expectation value of the Hamiltonian. Its relatively low circuit depth and partial robustness to noise make it particularly suitable for implementation on NISQ hardware.

VQE was introduced in the context of quantum chemistry \cite{Peruzzo_2014,McClean_2016}, and since then it has been applied to a wide range of problems, including quantum chemistry 
~\cite{O_Malley_2016, Kandala_2017, Parrish_2019, Arute_2020, Cao_2022, Zhao_2023}
and condensed matter systems ~\cite{Kokail_2019, Tilly_2022, Jattana_2022, Stanisic_2022}.
Extensions to lattice gauge theories \cite{Ciavarella_2022, Zhang_2023, Maiti_2025}, including the Schwinger model in 1+1 dimensions, have demonstrated the potential of variational quantum algorithms for simulating quantum field theories and, ultimately, real-time dynamics \cite{Nagano_2023, Notarnicola_2020}. Despite these promising developments, several challenges limit the scalability of VQE. The main obstacles among these are the emergence of barren plateaus \cite{McClean_2018} in the optimization landscape, the strong dependence on the choice of ans\"atz, the large measurement overhead arising from Hamiltonians with many Pauli terms, and the impact of noise and decoherence in deeper circuits \cite{Larocca_2025}. Recent benchmarking studies \cite{Wu_2024} have further highlighted that the optimization difficulty increases with the degree of correlation and connectivity in the system, posing additional challenges for higher-dimensional lattice models.

Understanding the interplay between ans\"atz expressivity, optimization stability, and entanglement is therefore crucial for assessing the viability of VQE in quantum many-body physics. In this paper, we address this issue by systematically studying the transverse-field Ising model (TFIM) in one, two, and three dimensions across different system sizes. The TFIM serves as an ideal testbed due to its well-understood quantum phase transition and the increasing complexity of its entanglement structure with dimensionality. Benchmarking VQE results for the TFIM against known solutions provides a means to optimize the VQE algorithm and its ansätze, thereby enabling the study of TFIM and similar systems  with significantly larger numbers of spins that are intractable using classical algorithms.  While VQE has previously been applied to one-dimensional chains and two-dimensional lattices \cite{kirmani2025variationalquantumsimulationstwodimensional, Sumeet_2024}, these studies have identified significant challenges, including sensitivity to ans\"atz choice and non-monotonic convergence with circuit depth, particularly near the critical point.

Here, we extend these investigations by providing a detailed benchmarking study of VQE performance across different spatial dimensions and, to the best of our knowledge, apply VQE to the TFIM on a three-dimensional cubic lattice of size $3 \times 3 \times 3$, corresponding to 27 spins. We focus on periodic boundary conditions, where the ground state is degenerate and highly entangled below the critical field, making the choice of ans\"atz an important consideration. To characterize ans\"atz performance, we employ measures of expressivity based on the Haar distribution of random unitaries, together with diagnostics for assessing the quality of the obtained ground state.

We assess VQE performance by comparing with exact diagonalization results and examining entanglement properties across different system sizes and geometries. In particular, we study how ans\"atz expressivity and optimization behavior affect the representation of highly entangled quantum states. We consider several ans\"atze, including the hardware-efficient EfficientSU2 implementation in Qiskit, the physics-inspired Hamiltonian Variational Ansatz (HVA), and its symmetry-breaking extension. Their performance is evaluated using energy variance, entanglement entropy, spin correlations, and magnetization. Overall, our results suggest a trade-off in VQE design: the hardware-efficient ansatz tends to exhibit smoother optimization landscapes but generally produces lower-fidelity ground-state approximations, whereas the physics-inspired ansatz tends to reproduce correlations more accurately and achieve higher fidelities, particularly in the strongly correlated regime below the critical field, albeit with a more challenging optimization landscape.

The paper is organized as follows: In Sec. \ref{sec:tfim}, for completeness, we briefly described the TFIM model. In Sec. \ref{sec:vqe} we briefly outline the VQE framework for this work. 
A discussion of expressivity and measurements is presented in Sec. \ref{sec:expr}, while the observables employed to characterize the properties of the TFIM are described in the subsequent subsections of Sec. \ref{sec:observ}. Simulation details are provided in Sec. \ref{sec:simu}, and results are presented in Sec. \ref{sec:results}. Finally, we summarize our findings and discuss future directions in Sec. \ref{sec:conclusion}.

\section{\label{sec:tfim}Transverse Field Ising Model}

The well-studied transverse field Ising model provides a minimal framework for understanding quantum phase transitions \cite{suzuki-2012, sachdev-2011}. In one dimension, the model is exactly solvable, while in higher dimensions it remains nontrivial and captures rich physics relevant to real materials. In this work, we investigate the TFIM in one, two, and three dimensions (with system size up to 27 spins), focusing on the interplay between ans\"atz expressivity and entanglement properties within the VQE framework. The Hamiltonian for the TFIM is defined as,
\begin{equation}
H = J_z \sum_{\langle ij \rangle} \sigma_i^z \sigma_j^z + h_x \sum_i \sigma_i^x,
\end{equation}
where $J_z$ is the nearest-neighbor interaction strength, and $h_x$ is the transverse magnetic field. We consider ferromagnetic interactions and set $J_z = -1.0$ while applying periodic boundary conditions throughout this work. The positive sign on the transverse-field is chosen to make the model non-stoquastic. 
The TFIM exhibits a quantum phase transition in the thermodynamic limit in one, two, and three dimensions at critical transverse fields $h_x^{(1D)}  = 1.0$, $h_x^{(2D)}  \approx 3.3$, and $h_x^{(3D)}  \approx 5.3$, respectively \cite{10.1007/978-3-642-83154-6_9,braiorrorrs2015phasediagramonetwo}. 

\section{\label{sec:vqe}Variational Quantum Eigensolver (VQE)}
VQE is a hybrid classical-quantum algorithm that can be employed to calculate the ground state of a many-body system by minimizing the energy expectation value of a variational quantum state $ |\psi(\vec{\theta}) \rangle$ \cite{Peruzzo_2014},
\begin{equation}
E_0 = \min_{\vec{\theta}} \bra{\psi(\vec{\theta})}  H \ket{\psi(\vec{\theta})}.
\end{equation}

VQE has two fundamental assumptions: 1. The given quantum circuit with its variational parameters can represent the ground state of the system. 2. Optimizer technique employed can successfully reach the desired parameters, i.e. the ground state. So, the problem also becomes twofold: (I) design the parametric quantum circuit (PQC), and (II) employ the optimizer. In Appendix   \ref{vqe_dis}, we provide a detailed discussion of VQE, the underlying assumptions, and their limitations, illustrated through a simple toy model.

\begin{figure}[htb]
    \centering
    \begin{subfigure}{0.99\linewidth}
        \includegraphics[width = \textwidth, height= 4.5cm]{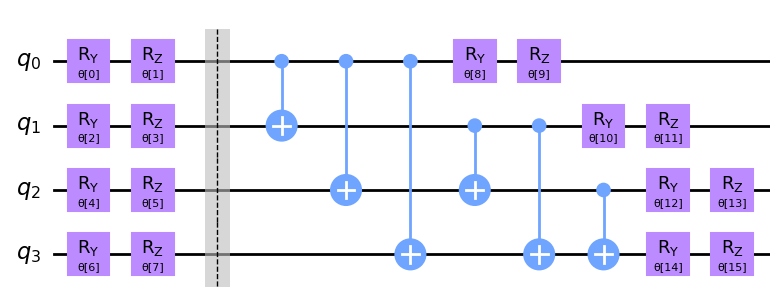}
        \caption{HEA}
        \label{fig:HEA_circuit}
    \end{subfigure}
    \begin{subfigure}{0.99\linewidth}
        \includegraphics[width = \textwidth, height= 4.5cm]{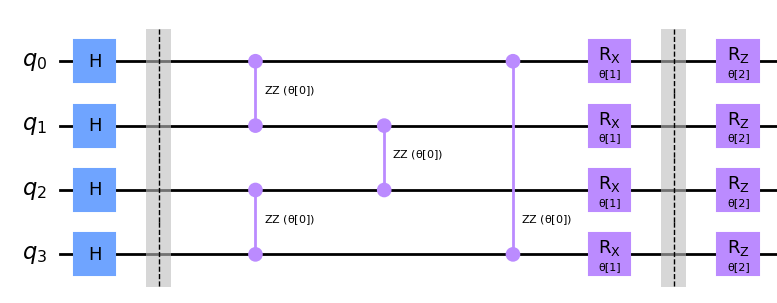}
        \caption{HVA with SB layer}
        \label{fig:HVA_SB_circuit}
    \end{subfigure}
    \caption{Schematic of the VQE circuits for HEA (hardware efficient ans\"atz), HVA (Hamiltonian variational ans\"atz) and HVA-SB (symmetry-breaking) used in this work.}
    \label{fig:HEA_HVA_circuits}
\end{figure}

To investigate the ans\"atz dependencies in our study, we used three different kinds of ans\"atz circuits: hardware-efficient ans\"atz (HEA), Hamiltonian variational ans\"atz (HVA), and HVA with a symmetry-breaking layer (HVA-SB). HEA is constructed using native hardware gates, making it easier to implement on a quantum processor. Qiskit's EfficientSU2 \cite{qiskit2024} ans\"atz is a well-known hardware-efficient ans\"atz (HEA). The HVA circuit \cite{Wiersema_2020} is produced by trotterization of the Hamiltonian into constituent non-commuting terms and applying them layer by layer in the circuit with variational parameters. It mimics the imaginary-time evolution of the Hamiltonian. HVA requires a smaller number of parameters due to the fact that all commuting terms share a single variational parameter. Symmetry-breaking layers \cite{park2024efficientgroundstatepreparation} can be integrated into the HVA ans\"atz to improve expressivity. In Fig. \ref{fig:HEA_HVA_circuits}, we schematically show the HEA and HVA-SB ans\"atze. The barriers are shown in the circuit as dotted lines in gray strips. 
In Fig.~\ref{fig:HEA_circuit}, the left side is the initial rotation, and the right side of the barrier is counted as one layer. For simulations, multiple layers are applied to reach the desired ground state.  The HVA-SB circuit is shown in Fig.~\ref{fig:HVA_SB_circuit}. The circuit is divided into three parts by barriers. The left part is the initial state, the middle part is the HVA layer, and the right part is the symmetry-breaking (SB) layer. For the HVA circuit, we do not use the SB layer; only the HVA layers are repeated after the initialization part. The SB layer is introduced only in the HVA-SB ans\"atz, where, for every layer, the HVA and SB layers are considered as a single layer and repeated accordingly. The resource estimation of the above ans\"atze in terms of the total number of CNOT gates, circuit depth, and number of parameters are discussed in Appendix \ref{vqe_resource}.

\section{\label{sec:observ} Observables for benchmarking VQE results}
In order to understand the properties of the ground state that VQE lead to, and to benchmark its accuracy, we employ various observables, namely expressivity, energy variance, spin correlations and entanglement entropy. We discuss those below.

\subsection{\label{sec:expr} Expressivity}
A circuit's expressivity is described by its effectiveness in producing uniformly distributed random states in a given Hilbert space
by random parameter sets. The expressivity of an ans\"atz can be quantified as the Haar measure of random unitary which is called the frame potential. The frame potential for Haar-random pure state \cite{Roberts_2017} in dimension $d$ for $t$-design is defined as the mean of the $t$-momentum or $t$-degree norm of the two set of randomly generated states. Mathematically, the frame potential can be defined as \cite{Sim_2019}
\begin{equation}
    F_t = {\mathbb{E}}_{\psi,\phi} \left[ \left| \langle \psi | \phi \rangle \right|^{ 2t } \right],
    \label{eq:Haar_measure}
\end{equation}
where the states $\ket{\psi}$ and $\ket{\phi}$ are generated from two random parameter sets, and $\mathbb{E}$ denotes the mean over the entire sample of these states. The $t$-design or $t$-momentum means that the state is randomized to a $t$-degree of application. Since we are interested primarily in energy calculation, we will use a 1-degree application for our study.
\begin{figure}[htb]
    \centering
\includegraphics[width=\linewidth]{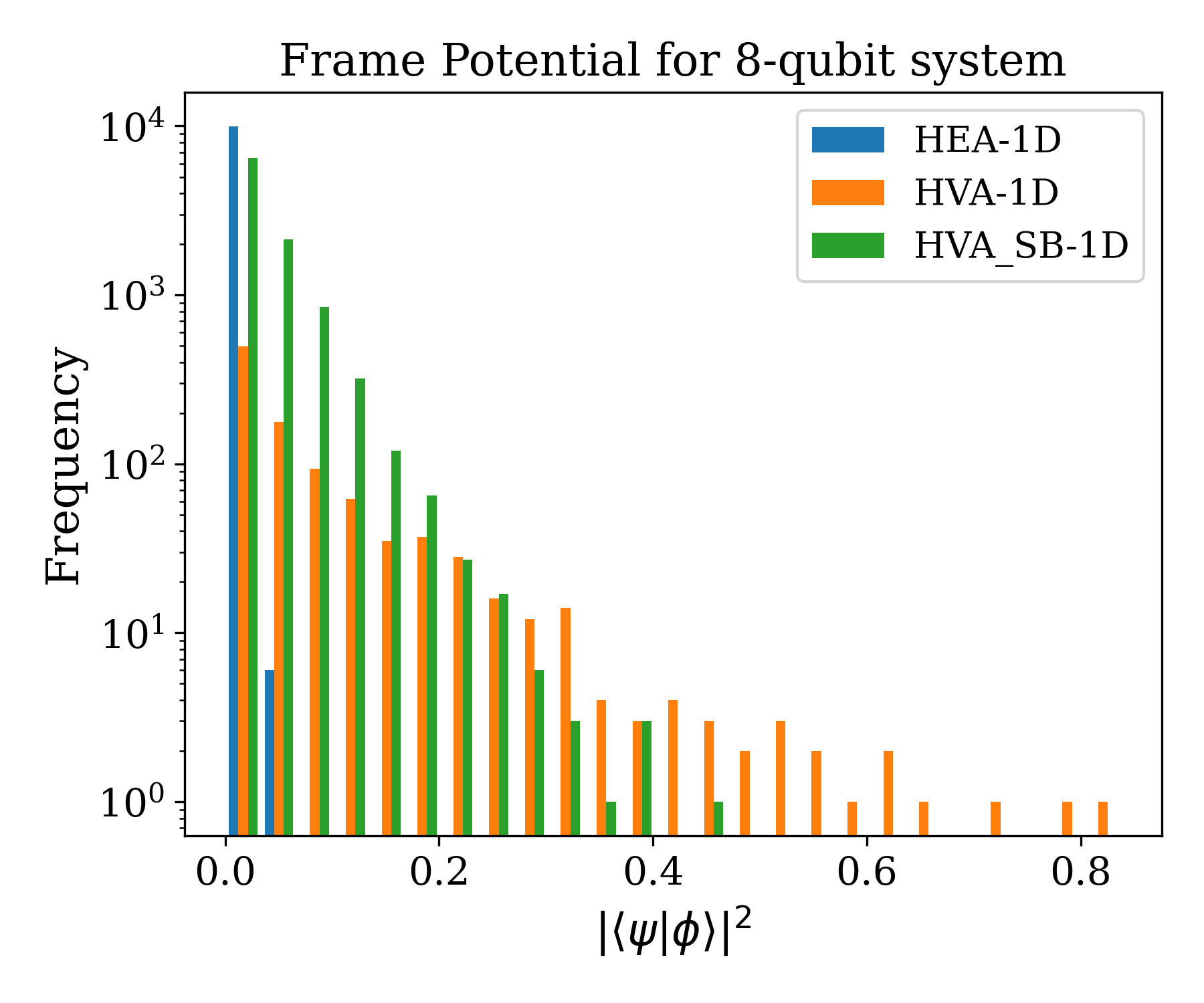}
    \caption{{Distribution of 1-degree norm for HEA, HVA and HVA-SB ans\"atze for pairs of states generated respectively from two random parameter sets. The frame potential $F_1$ -- the average of the 1-degree norm over random parameter sets -- is a measure of expressivity. The Hamiltonian is 1D TFIM.}}
    \label{fig:frame_potential}
\end{figure}

A small value of the frame potential indicates that the states are more uniformly distributed, corresponding to higher expressivity. Highly expressive ans\"atz effectively explore a nearly uniform subspace of the full Hilbert space, whereas less expressive circuits are restricted to a smaller, more specific subspace. 
We have calculated the 1-design frame potential for all three ans\"atz circuits that we employed. The comparative effectiveness 
for the HEA, HVA, and HVA-SB ans\"atz circuits are shown in 
Fig. \ref{eq:Haar_measure} with a histogram plot. We have taken 10000 samples of states $\ket{\phi}$ and $\ket{\psi}$ generated from two random sets of parameters and plotted their frequency vs their norm value $\left| \langle \psi | \phi \rangle \right|^{2}$ . 
From this, we observe that the HEA exhibits the highest expressivity, the HVA the lowest, and the HVA-SB ans\"atz shows enhanced expressivity, as expected due to the inclusion of the symmetry-breaking layer. Note that the frame potential is a good indicator of how well an ans\"atz can approximate the desired ground state, rather than actually finding the ground state. It serves the purpose of reaching the approximate ground state energy with a highly expressive circuit, after which we can apply an ans\"atz in the restricted Hilbert space of the given Hamiltonian to find the exact ground state. 

\subsection{Energy variance}
This takes us to the next quantity in our benchmarking process: the energy variance. In our calculation, we first increase the number of layers until we get a stable value for the energy; then we use energy variance to find the precision of the ground state. We define it as,
\begin{equation}
    Var(E) = \frac{\langle H^2 \rangle - \langle H \rangle^2}{\langle H \rangle^2}   
    \label{eq:var_E},
\end{equation}
and it is linearly dependent on the error in the eigenenergy \cite{Wu_2024}. However, it must be kept in mind that one needs to  be close to the ground state because energy variance can also vanish for any of the eigenstates of the system. Once it reaches closer to the ground state of the system, the energy variance helps to compute it with better precision.

\subsection{\label{sec:spin_corr}Spin correlations}
The TFIM exhibits a quantum phase transition in the thermodynamic limit. However, for a finite system, $\mathbb{Z}_2$ symmetry is not spontaneously broken: in the absence of the transverse field, one gets two low-lying eigenstates with a small energy gap proportional to $e^{-N}$, where $N$ is the system size. These nearly degenerate states conserve parity. Denoting these states as $\ket{\psi_0}$ and $\ket{\psi_1}$, and considering the parity preservation in a finite TFIM system, one can express for small but nonzero transverse fields,
\begin{equation}
    \begin{split}
        \ket{\psi_0} &= c_0(\ket{00..00} + \ket{11..11}) + .... , \\
        \ket{\psi_1} &= c_0(\ket{00..00} - \ket{11..11}) + ....
    \end{split}
\end{equation}
 
In this case the magnetization, 
\begin{equation}
    M = \frac{1}{N} \sum_i \langle \sigma^z_i \rangle ,
\end{equation}
is always zero for the ground state. However, one can still get a net magnetization in VQE in the low-field regime due to the superposition of the energetically close ground and first excited states, and hence it is not a good probe to study the ground state properties.
For probing the ground state properties, one can instead employ the spin correlation function, defined as,
\begin{equation}
    M_{Corr} = \frac{1}{N/2} \sum_i \langle \sigma^z_i \sigma^z_{(i+N/2)} \rangle,~~~~~~~~\text{for even N},
\end{equation}
{Although the spin correlation function is standard in conventional studies of the TFIM, its use for benchmarking VQE approach remains relatively unexplored, and in this work we have utilized it.}

\subsection{Entanglement Entropy}

Entanglement plays a central role in quantum simulations. We compute the entanglement entropy of the ground state to assess the required quantum resources, as well as to understand the role of ans\"atz expressivity and circuit design in accurately reaching the true ground state. As discussed earlier, for a finite system of the TFIM, the $\mathbb{Z}_2$ symmetry does not get spontaneously broken below the critical field. Instead, the system exhibits a highly entangled, nearly degenerate ground state. Hence, by evaluating the bipartite entanglement entropy, we can determine whether a given ans\"atz or VQE state is capable of faithfully capturing this entanglement structure. 
The bipartite entanglement entropy is obtained in the usual manner:  for a state $\ket{\psi}$, we trace over the partition $B$ and obtain the reduced density matrix for its complement (here, for partition A),
\begin{equation}
\begin{aligned}
\rho_A &= \mathrm{Tr}_{B}(\rho)  \\
&= \sum_{\phi_B} \bra{\phi_B} \left( \ket{\psi}\bra{\psi} \right) \ket{\phi_B} \\
&= \sum_{\phi_B} \bra{\psi}\ket{\phi_B}\bra{\phi_B}\ket{\psi}, \\
\text{and}\qquad
\rho_{A(i,j)} &= \bra{\psi}\ket{\phi_A}_i \bra{\phi_A}_j\ket{\psi}.
\end{aligned}
\label{eq:red_density_matrix}
\end{equation}
{In particular,} the partial trace for a one-qubit subsystem (for $k^{\text{th}}$ - qubit) is
\begin{equation}
    \rho_k = 
    \bra{\psi}
    \begin{bmatrix}
        \ket{0}^k \bra{0}^k & \ket{0}^k \bra{1}^k \\
        \ket{1}^k \bra{0}^k & \ket{1}^k \bra{1}^k
    \end{bmatrix}
    \ket{\psi},
\end{equation}
where,
\begin{equation}
    \begin{split}   
        \ket{0} \bra{0} & = \frac{1}{2} \left( I + \sigma_z \right),~~~~
        \ket{0} \bra{1} = \frac{1}{2} \left( \sigma_x - \iota \sigma_y \right), \\        
        \ket{1} \bra{1} & = \frac{1}{2} \left( I - \sigma_z \right),~~~~ \ket{1} \bra{0} = \frac{1}{2} \left( \sigma_x + \iota \sigma_y \right).       
    \end{split}
\end{equation}
The entanglement entropy is given by
\begin{equation}
        S_{EE}^{BE} = -\sum_i\lambda_i \log{\lambda_i},
        \label{eq:entropy}
\end{equation}
where $\{\lambda_i\}$ are the eigenvalues of $\rho_k$.

Alternatively, one can also use singular value decomposition to calculate the entanglement entropy 
for the ground state obtained in a VQE setup using the state tomography method. A pure state, $\ket{\psi}$, in a Hilbert space, $H = A \otimes B$,  can be expressed as a Schmidt decomposition as,
\begin{equation}
    \ket{\psi} = \sum_i^m \alpha_i \ket{\phi_A}_i \otimes \ket{\phi_B}_i,
\end{equation}
where, $m$ is the minimum of the total number of basis states in $ \{\ket{\phi_A} \} $ and $ \{\ket{\phi_B}\} $. The entanglement entropy is then given by
\begin{equation}
    S_{EE}^{SVD} = - \sum_i \alpha_i^2 \log{\alpha_i^2}.
    \label{eq:SVD}
\end{equation}
Computation of the entanglement entropy using VQE is possible for small subsystem sizes, and the two schemes (Eqs.  \ref{eq:entropy} and \ref{eq:SVD}) provide complementary insights into entanglement calculations by directly accessing the reduced density matrix and von Neumann entropy. That is an advantage over Quantum Monte Carlo methods (QMC). In a Quantum Monte Carlo (QMC) calculation, the absence of direct access to the eigenstates or the reduced density matrix prevents a direct evaluation of the von Neumann entropy. Instead, one can compute R\'enyi entropies of integer order $n \ge 2$, most commonly the second-order R\'enyi entropy, using replica-based techniques~\cite{Hastings_2010}. Although several numerical approaches have been attempted to estimate the von Neumann entropy by extrapolating these higher-order R\'enyi entropies to the $n \to 1$ limit, for instance via reconstruction of the entanglement spectrum~\cite{Dalmonte_2018, Mendes_Santos_2020}, such procedures are indirect and rely on approximations rather than a direct computation.
Note that, in a VQE setup, the R\'enyi entropy can also be computed using alternative approaches, such as doubling the number of qubits or employing statistical techniques like classical shadows. However, the method proposed above using Eq. (\ref{eq:entropy}) allows direct calculation of the von Neumann entropy from the reduced density matrix of a small subsystem, while Eq. (\ref{eq:SVD}) relies on full state tomography of the quantum state.

\section{\label{sec:simu}Simulation Details}
We implement VQE using the CUDA-Q framework in state-vector simulation mode, which leverages the NVIDIA cuQuantum libraries \cite{the_cuquantum_development_team_2023_10068206} to accelerate large-scale quantum simulations on GPUs. This allows us to treat 1D, 2D, and 3D TFIM systems for larger lattice sizes, while still benchmarking ans\"atz performance and optimization strategies. The exact simulation of the VQE without any statistical noise shows us how our method will perform in an ideal condition. 

First, we adopt Qiskit's EfficientSU2 ans\"atz (HEA ans\"atz, as shown in Fig. \ref{fig:HEA_circuit}) in the CUDA-Q framework. We restrict it to only real amplitudes by removing the $R_z$-gates from the circuit. This helps us reduce the number of parameters while maintaining high expressivity. This is our first step in the VQE simulation. We use this ans\"atz to calculate the energy as well as the entanglement entropy of multiple lattices in 1D, 2D, and 3D TFIM; utilizing up to 27 qubits. It allows us to determine the scaling challenges and highlights the question of the role of ans\"atz design and optimizers in VQE. This leads us to the exploration of different ans\"atz circuits and benchmarking them through the observables we have introduced above.

Benchmarking simulations are also carried out within the CUDA-Q framework. We employ three families of circuits: the hardware-efficient ans\"atz (HEA), the Hamiltonian variational ans\"atz (HVA), and a symmetry-breaking (SB) ans\"atz, as discussed in the Sec. \ref{sec:vqe}. The HEA is constructed from layers of single-qubit rotations and CNOT gates with all-to-all site entanglement. Because this ans\"atz produces a relatively smooth parameter landscape, we employ the L-BFGS optimizer, a quasi-Newton method that leverages curvature information and converges efficiently in smooth cost surfaces. In contrast, the HVA is tailored to the structure of the TFIM Hamiltonian, alternating between interaction and field terms within each variational layer. The SB ans\"atz explicitly allows for breaking of the $\mathbb{Z}_2$ parity symmetry by introducing an extra layer of $R_z$ gates beside the $R_x$ gates in the HVA circuit which increases the circuit depth but allows better energy optimization. Both HVA and HVA-SB produce more rugged optimization landscapes, so we employ COBYLA, a derivative-free constrained optimization method that is more robust in such cases. Both L-BFGS and COBYLA optimizers used here are inbuilt in the CUDA-Q python library.
For each ans\"atz, the number of layers is systematically varied, with 4, 8, 10, and 15 layers considered for both 1D and 2D TFIM simulations. This allows us to probe the trade-off between circuit expressivity and optimization difficulty.
From the optimized states, we compute physical observables such as magnetization, spin correlations, energy variance, and von Neumann entropy. This systematic comparison of ans\"atz families and optimizers highlights how the structure of the cost landscape strongly influences the choice of classical optimization method, and in turn, the efficiency of the overall VQE algorithm.

The benchmarking in VQE as a stand-alone approach comes from the fact that we can first explore the problem with a highly expressive ans\"atz which, in turn, gives us a close approximation of the ground state energy. Then we can employ the HVA, which by its construct is restricted to the Hamiltonian subspace providing us with the eigenstate of the system. Note that this ans\"atz is not optimizer-friendly and should be checked with different initial conditions for a reliable result.

\section{\label{sec:results}Results}
\begin{figure}[htb]
    \centering
    \begin{subfigure}{\linewidth}
        \centering 
        \includegraphics[width=\linewidth]{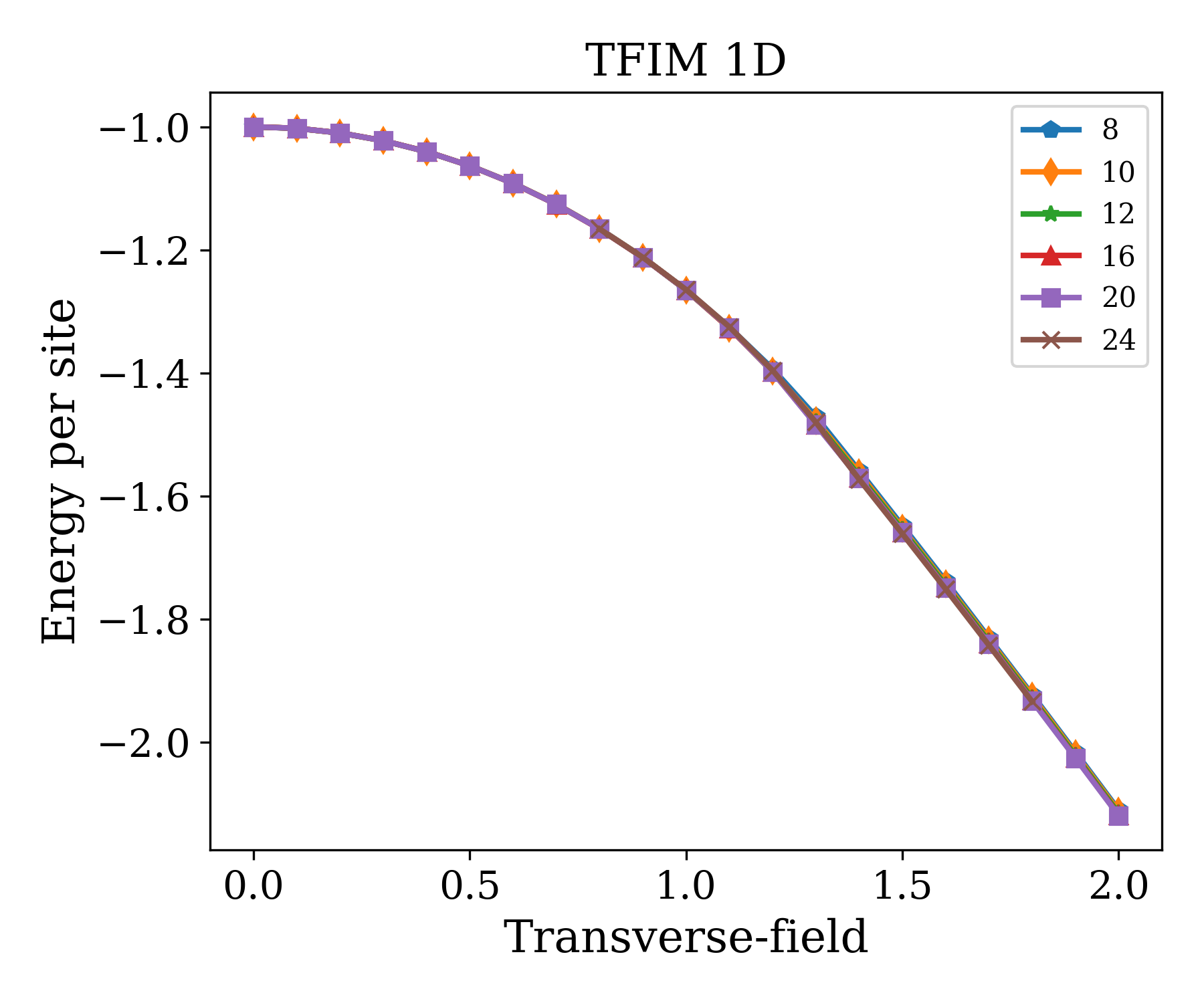}
        \caption{Average energy (energy per site)}
        \label{fig:1_d_Energy}
    \end{subfigure}
    \begin{subfigure}{\linewidth}
        \includegraphics[width=\linewidth]{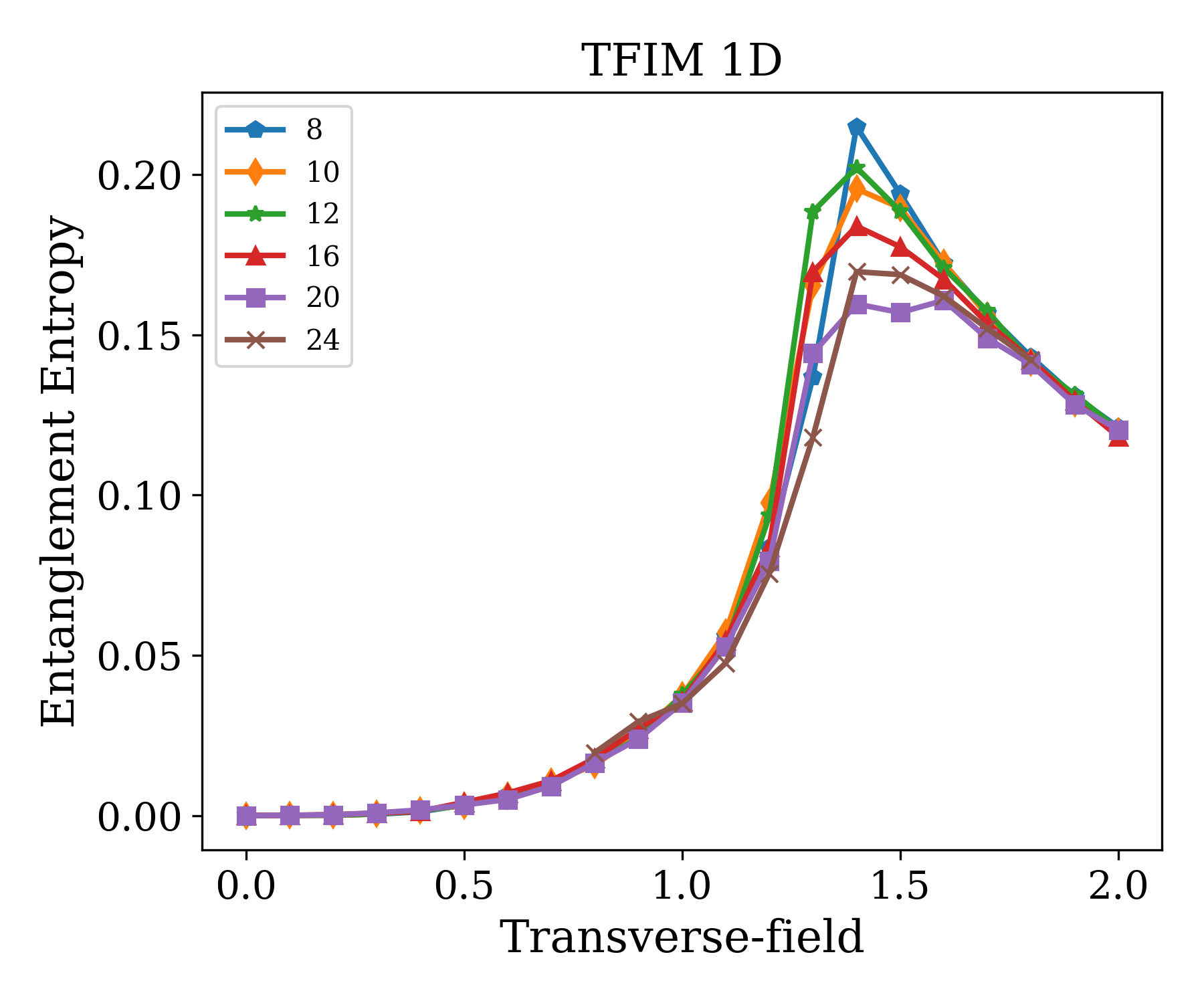}
        \caption{Entropy (single-site subsystem)}
        \label{fig:1_d_EE}
    \end{subfigure}
    \caption{Average energy and the von Neumann entanglement entropy for the ground state of the one dimensional TFIM at different lattice sizes.}
\end{figure}

In accordance with our workflow we first use HEA with Real Amplitude (HEA-RA) ans\"atz to compute the ground state properties of the TFIM in 1D, 2D and 3D spin clusters. 
The optimization time is reduced by using the optimized (ground state) parameters from the nearest transverse field value. HEA-RA is still a highly expressive circuit -- this can be easily seen by calculating the frame potential. We first calculate the energy for different lattice sizes in 1D. The results are presented in Fig.~\ref{fig:1_d_Energy} for different system sizes. The plot shows that the VQE-computed energies are in good agreement across system sizes. In Fig.~\ref{fig:1_d_EE}, we present the von Neumann entanglement entropy calculated using Eq.~\ref{eq:entropy}, which exhibits a clear peak near the critical point. 

\begin{figure}[htb]
    \centering
    \begin{subfigure}{\linewidth}
        \includegraphics[width=\linewidth]{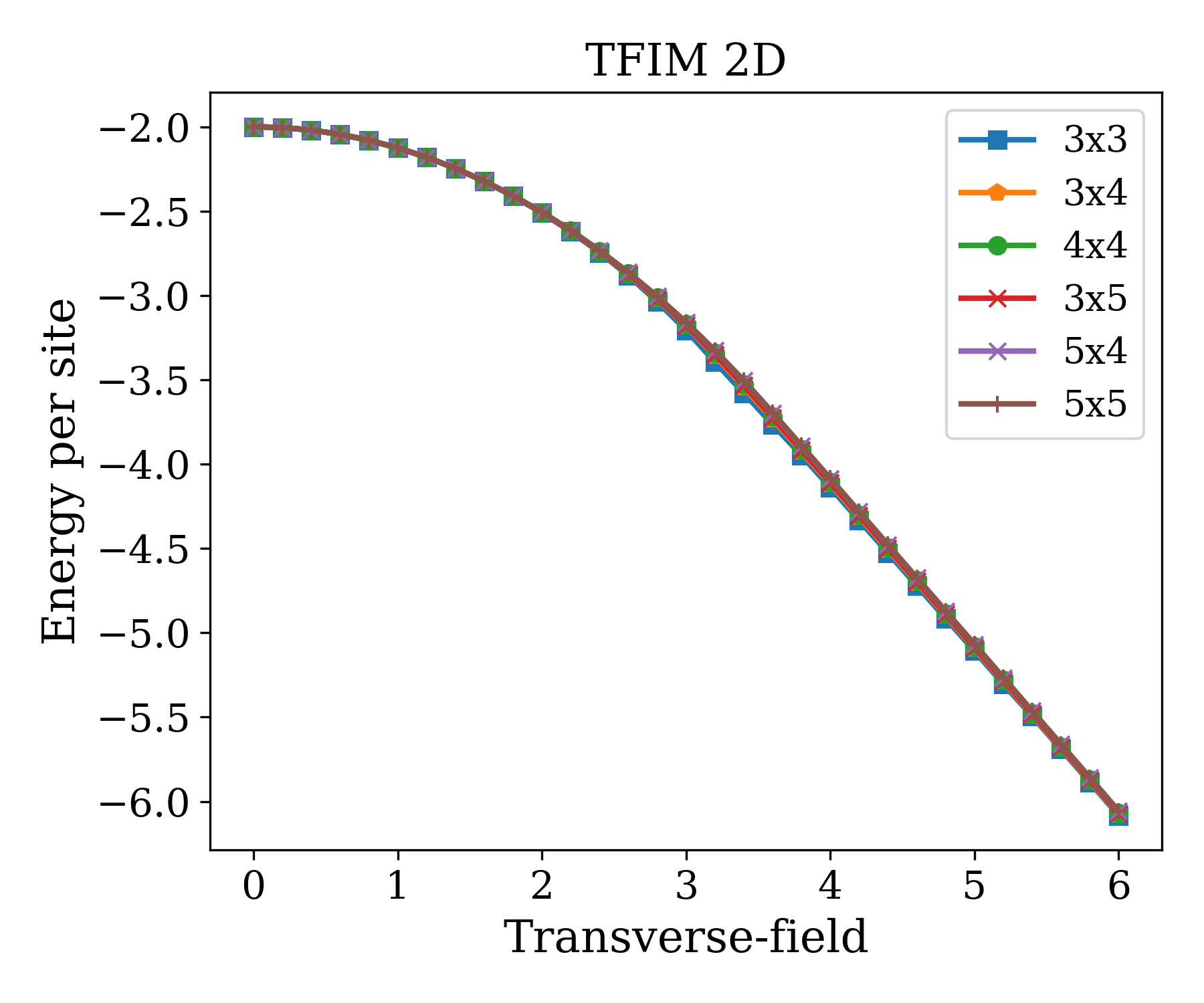}
        \caption{Average energy (energy per site)}
        \label{fig:2_d_Energy}
    \end{subfigure}
    \begin{subfigure}{\linewidth}
        \begin{overpic}[width=\linewidth]{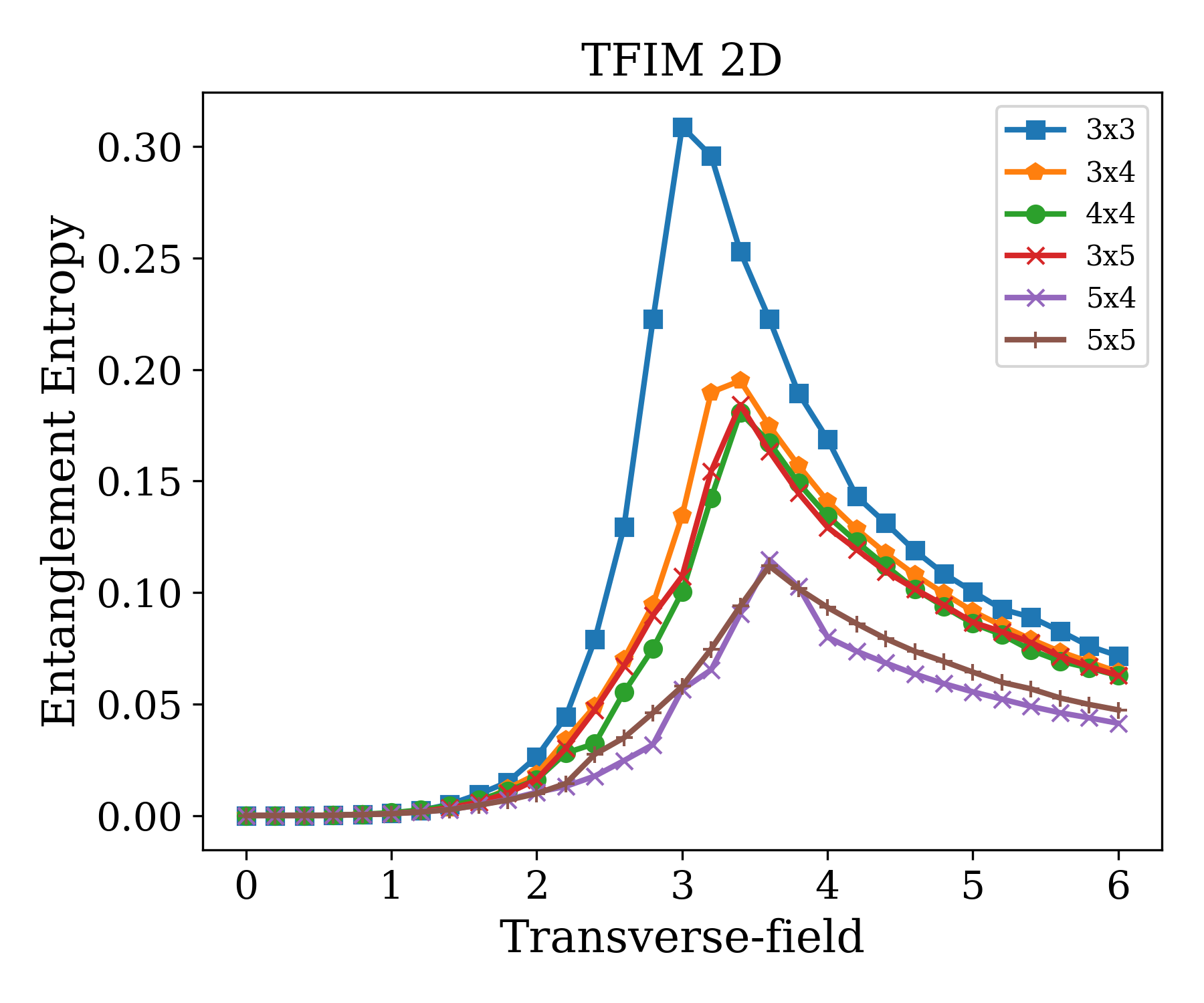}
            \put(18,47){\includegraphics[width=0.35\textwidth]{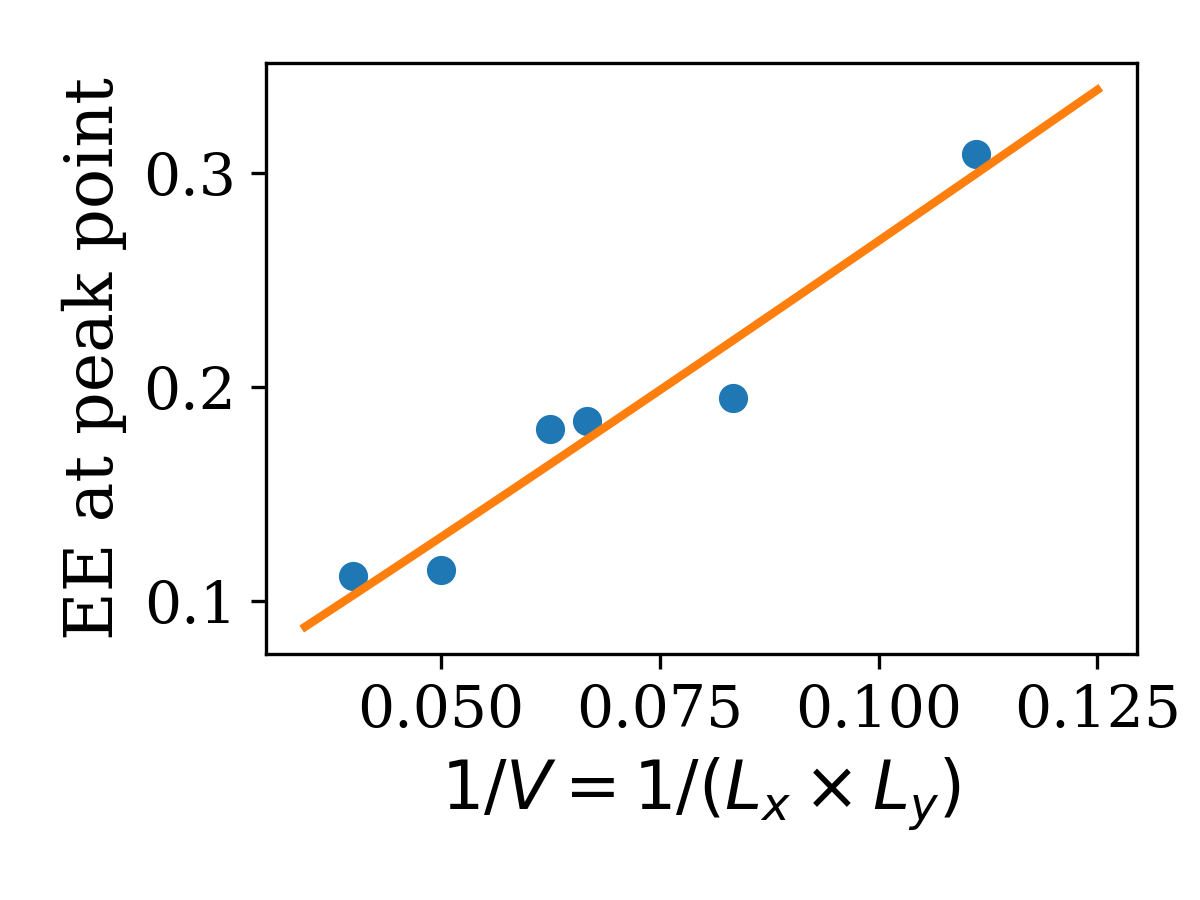}}
        \end{overpic}
        \caption{Entropy (single-site subsystem)}
        \label{fig:2_d_EE}
    \end{subfigure}
    \caption{Average energy and the von Neumann entanglement entropy for the ground state of the two dimensional TFIM at different lattice sizes.}
\end{figure}

In Figs.~\ref{fig:2_d_Energy} and \ref{fig:2_d_EE}, we present the results for the two-dimensional case. As in the one-dimensional case, the computed energy per site remains largely independent of system size (Fig.~\ref{fig:2_d_Energy}), and the entanglement entropy exhibits a peak near the critical point (Fig.~\ref{fig:2_d_EE}). However, with increasing dimensionality, we observe a reduction in the entanglement entropy as the system size grows. This can be understood from the fact that the entropy associated with a given qubit is effectively distributed among a larger number of neighboring qubits in higher dimensions, leading to a reduced entropy per site as the system size increases.

The optimization problem becomes more challenging in three dimensions; nevertheless, we are still able to compute the ground-state energies with reasonable accuracy using the HEA-RA ans\"atz. The corresponding energy and entanglement entropy results are shown in Figs.~\ref{fig:3_d_Energy} and \ref{fig:3_d_EE}, respectively. As in lower dimensions, the energy per site remains approximately independent of system size, indicating consistent ground-state estimates. The entanglement entropy exhibits signatures of critical behavior, although its magnitude decreases for larger system sizes, reflecting reduced entanglement per site in higher dimensions. {A quantitative analysis for the 3D system is not feasible with the current data, as a lattice of linear size 2 under periodic boundary conditions leads to each qubit sharing identical neighbors on both sides, hence limiting meaningful spatial distinctions.

\begin{figure}[htb]
    \centering
    \begin{subfigure}{\linewidth}
        \includegraphics[width=\linewidth]{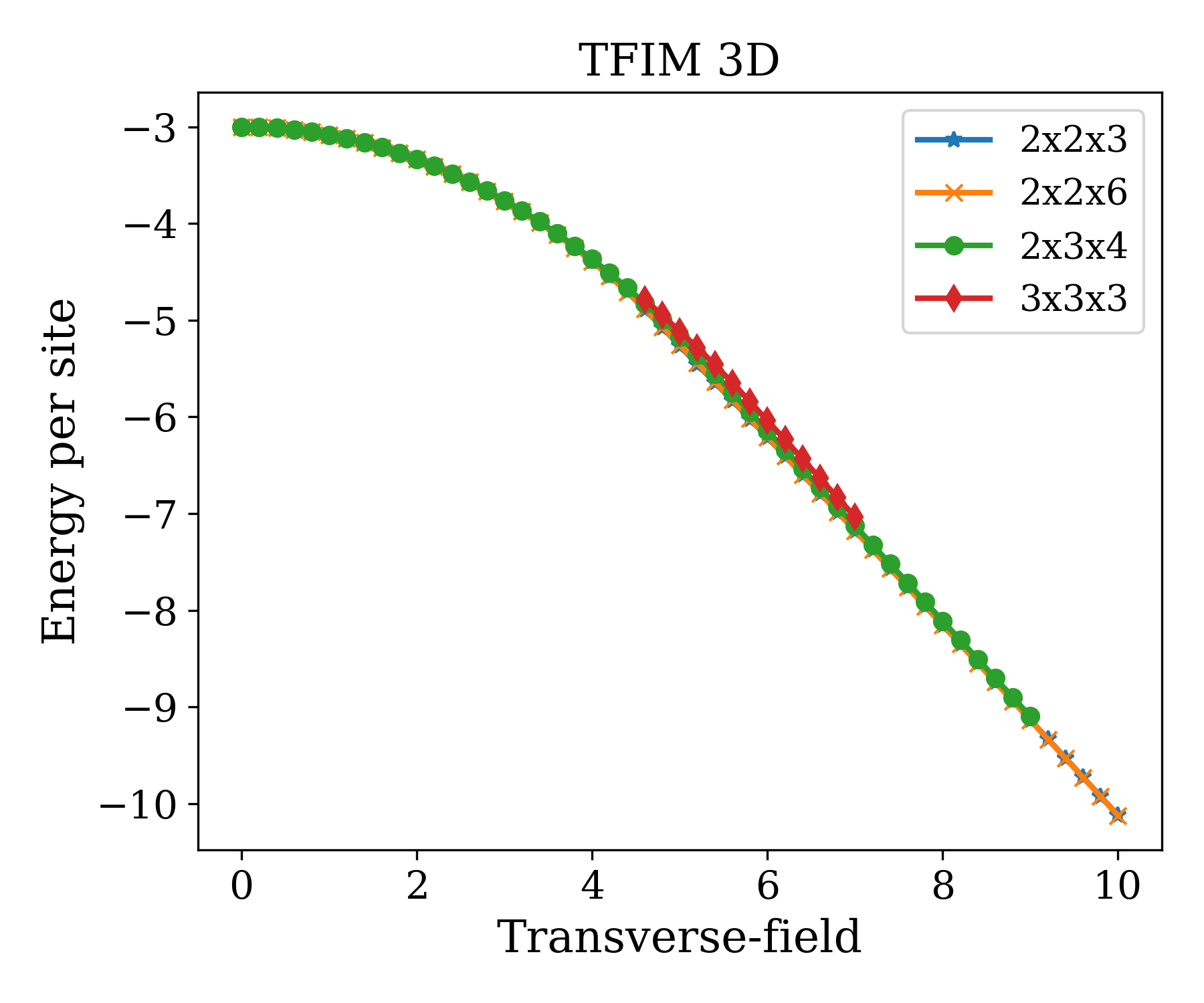}
        \caption{Average energy (energy per site)}
        \label{fig:3_d_Energy}
    \end{subfigure}
    \begin{subfigure}{\linewidth}
        \includegraphics[width=\linewidth]{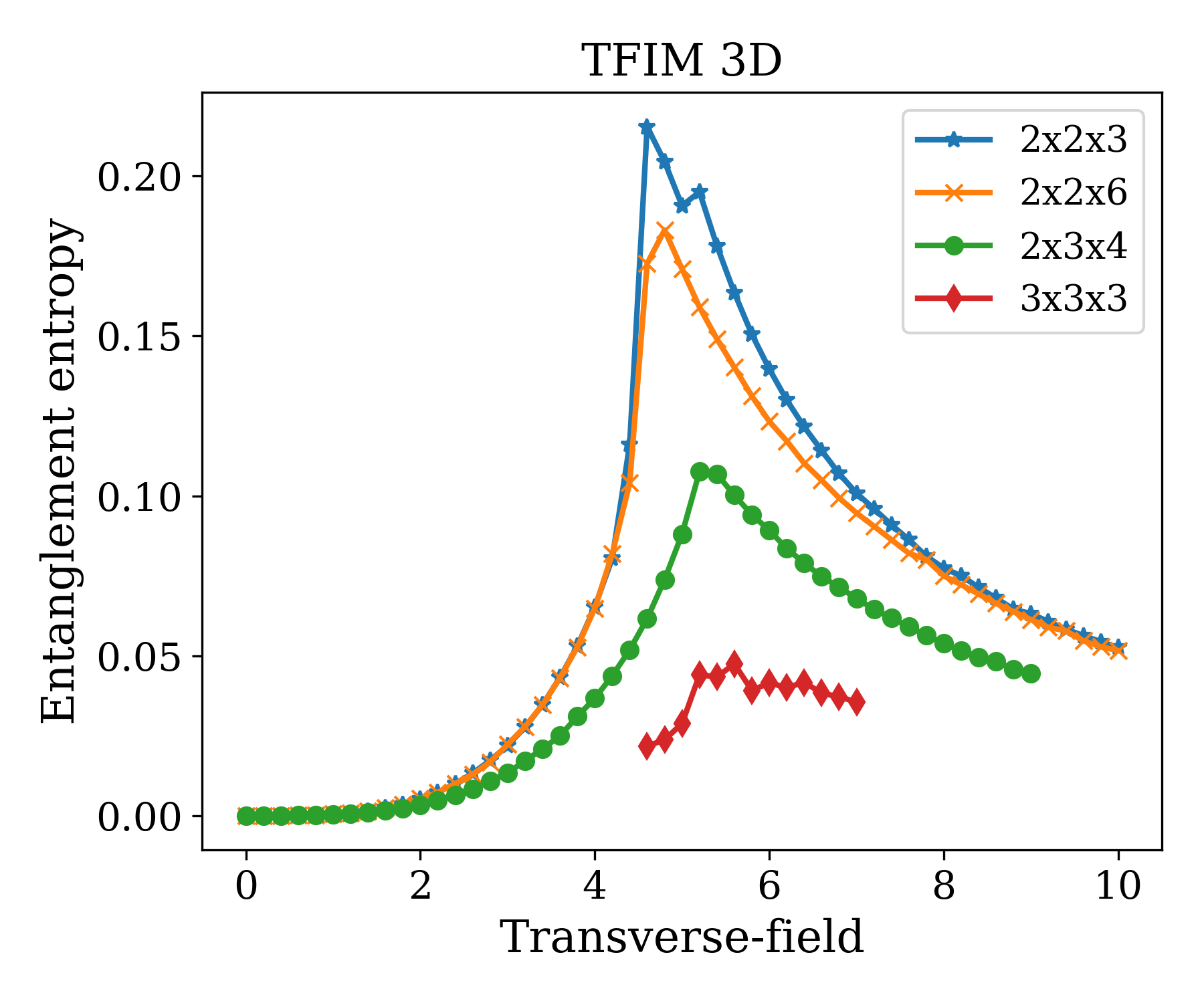}
        \caption{Entropy (single-site subsystem)}
        \label{fig:3_d_EE}
    \end{subfigure}
    \caption{Average energy and the von Neumann entanglement entropy for the ground state of the three dimensional TFIM at different lattice sizes.}
\end{figure}

\subsection{Benchmarking}

In this section we benchmark the VQE results against exact diagonalization and density functional renormalization group (DMRG). Numerical exact diagonalization (ED) has been the most commonly used method for computing the eigenvalues and eigenvectors of quantum systems. However, ED is severely limited by system size and quickly becomes intractable as the Hilbert space grows. In recent years, tensor network methods have also become increasingly popular, particularly for one-dimensional systems. Their limitations, however, appear in systems with large entanglement and in higher dimensions. VQE offers a hybrid classical-quantum approach in which, in principle, ground-state properties of larger systems can be studied. Below we compare these three methods. First, we benchmark VQE results against ED and DMRG (MPS), and then proceed to investigate the ans\"atz choices in VQE calculations.

In Fig.~\ref{fig:1_d_ED_DMRG_VQE}, we present a comparison of results obtained using ED, DMRG, and VQE for a one-dimensional TFIM system with 10 sites (details are given in Appendix  \ref{ED_DMRG_VQE_compare}). Here the observables are the ground state energy and the entanglement entropy. It is interesting to observe that while all methods can compute the energy accurately, to a certain extent, the VQE with the HVA ans\"atz shows better agreement with ED for the entanglement entropy (see Table \ref{tab:1D_ED_DMRG_VQE} in Appendix \ref{ED_DMRG_VQE_compare}). This can be attributed to the limitations discussed for DMRG, where we use a matrix product state (MPS) representation. We further observe that VQE does not accurately reproduce the ground state for the HEA and HVA-SB ans\"atze. This can be understood by noting that, although these ans\"atze are more expressive, they may fail to capture the correct entanglement structure 
and instead approximate a superposition of the two lowest nearly degenerate states.  
\begin{figure}[htb]
    \centering
    \begin{subfigure}{\linewidth}
        \includegraphics[width = \textwidth]{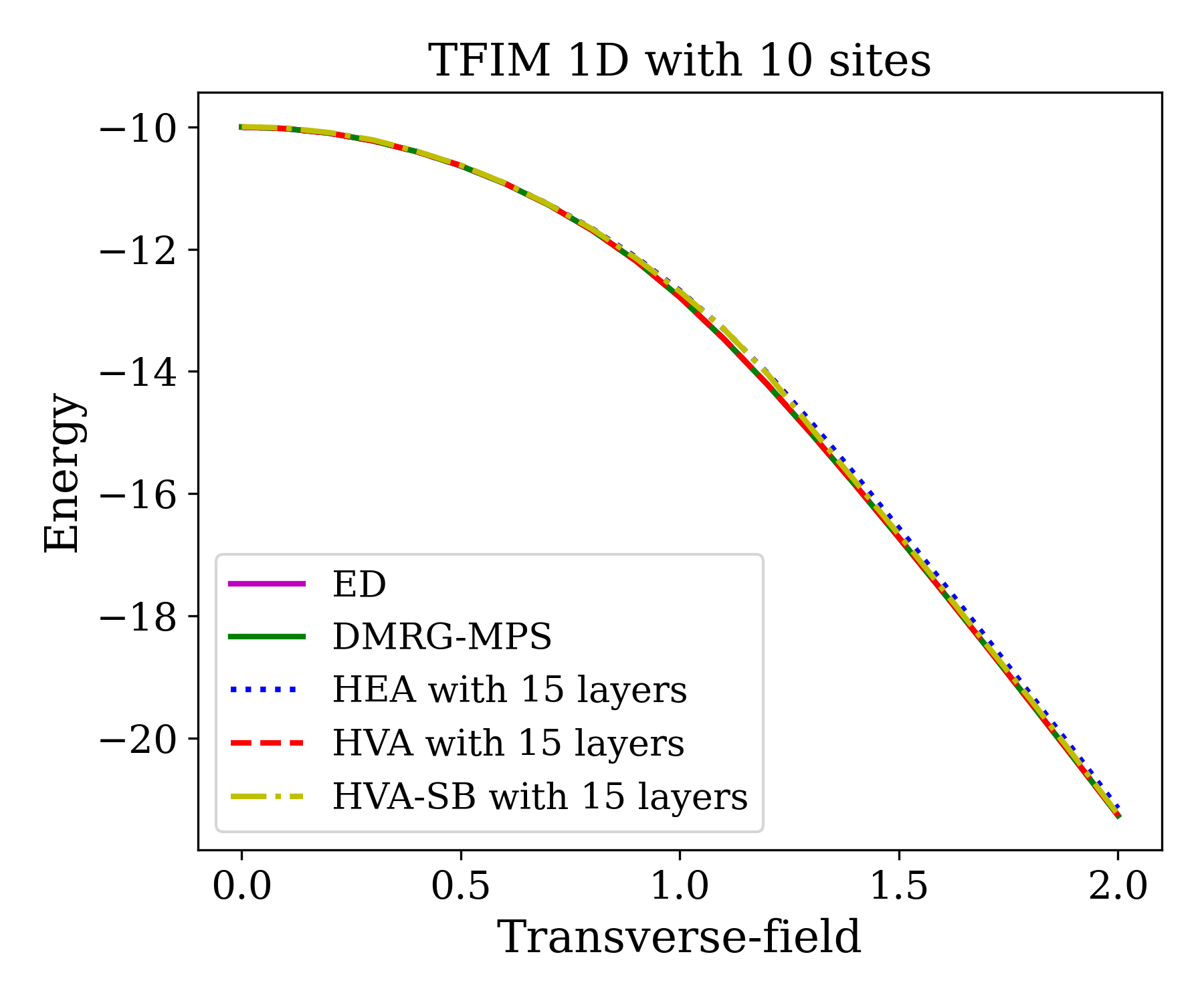}
        \caption{Energy}
        \label{fig:1_d_energy_10}
    \end{subfigure}
    \begin{subfigure}{\linewidth}
        \includegraphics[width = \textwidth]{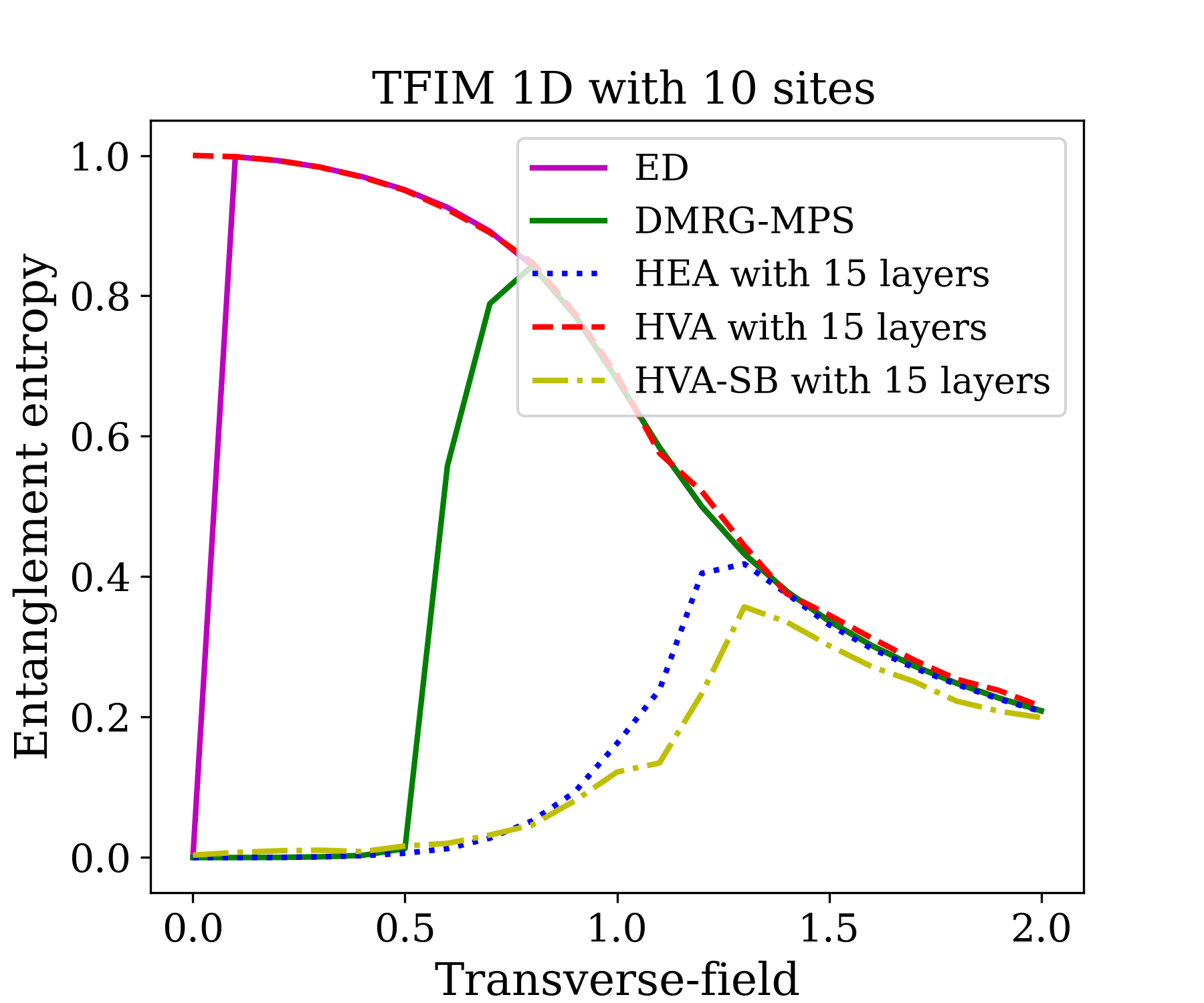}
        \caption{Entanglement entropy}
        \label{fig:1_d_EE_10}
    \end{subfigure}
    \caption{A comparison between ED, DMRG and VQE methods for energy and entanglement entropy}
    \label{fig:1_d_ED_DMRG_VQE}
\end{figure}

Next we investigate the ans\"atz choices for the results corresponding to observables discussed in Sec.~\ref{sec:observ}. We begin with the one-dimensional case. As demonstrated in earlier studies \cite{Sumeet_2024, kirmani2025variationalquantumsimulationstwodimensional}, and also in this work, the VQE framework performs very well in obtaining the ground state energy of the one-dimensional TFIM. Since the model is exactly solvable in one dimension, it provides an ideal tool for a comparative benchmarking of different observables across various ans\"atze. As discussed previously, we employ multiple ans\"atz layers and observe that the required number of layers grows polynomially with the number of sites in both one- and two-dimensional systems.

\begin{figure}[htb]
    \centering
    \includegraphics[width=\linewidth]{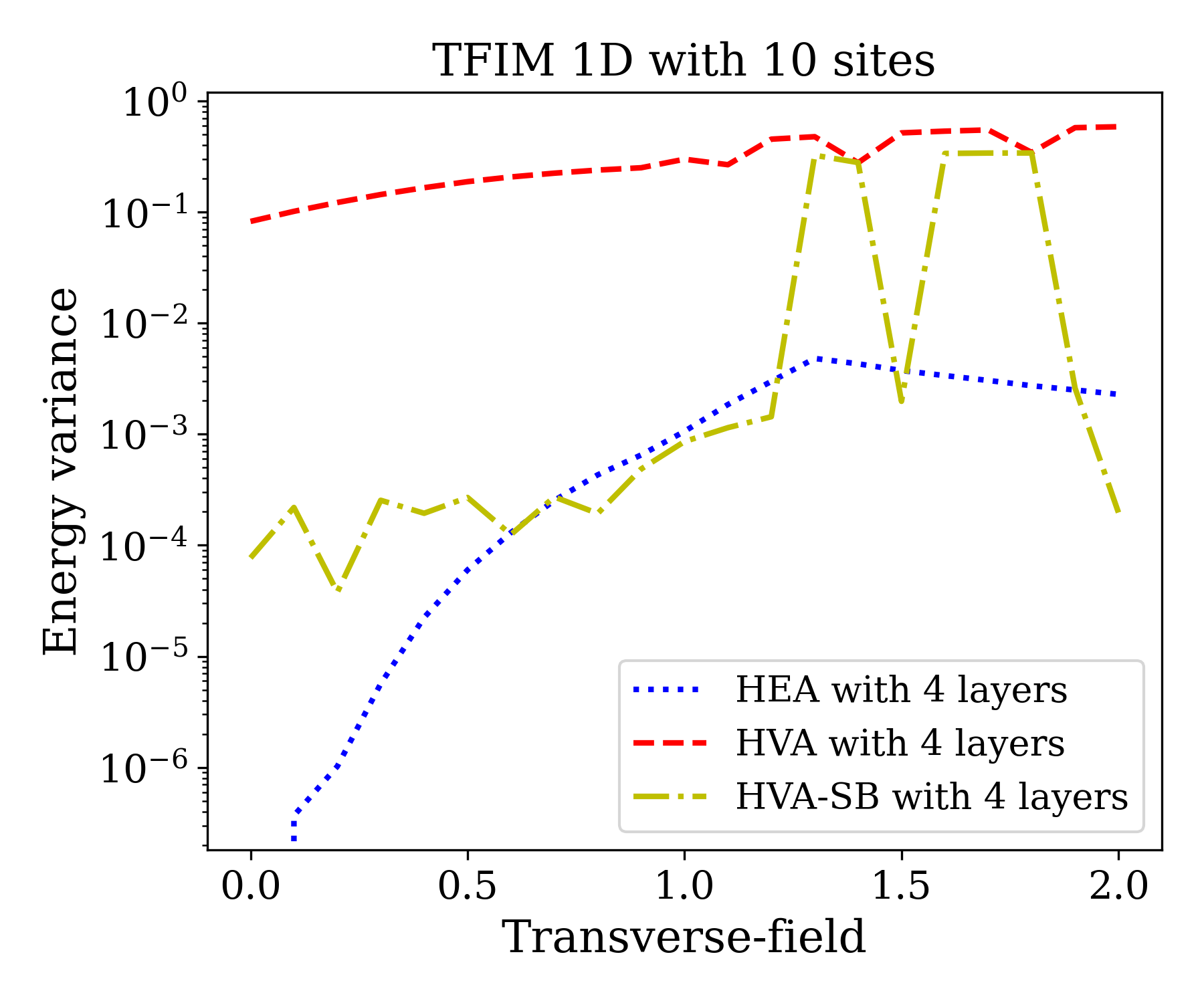}
    \includegraphics[width=\linewidth]{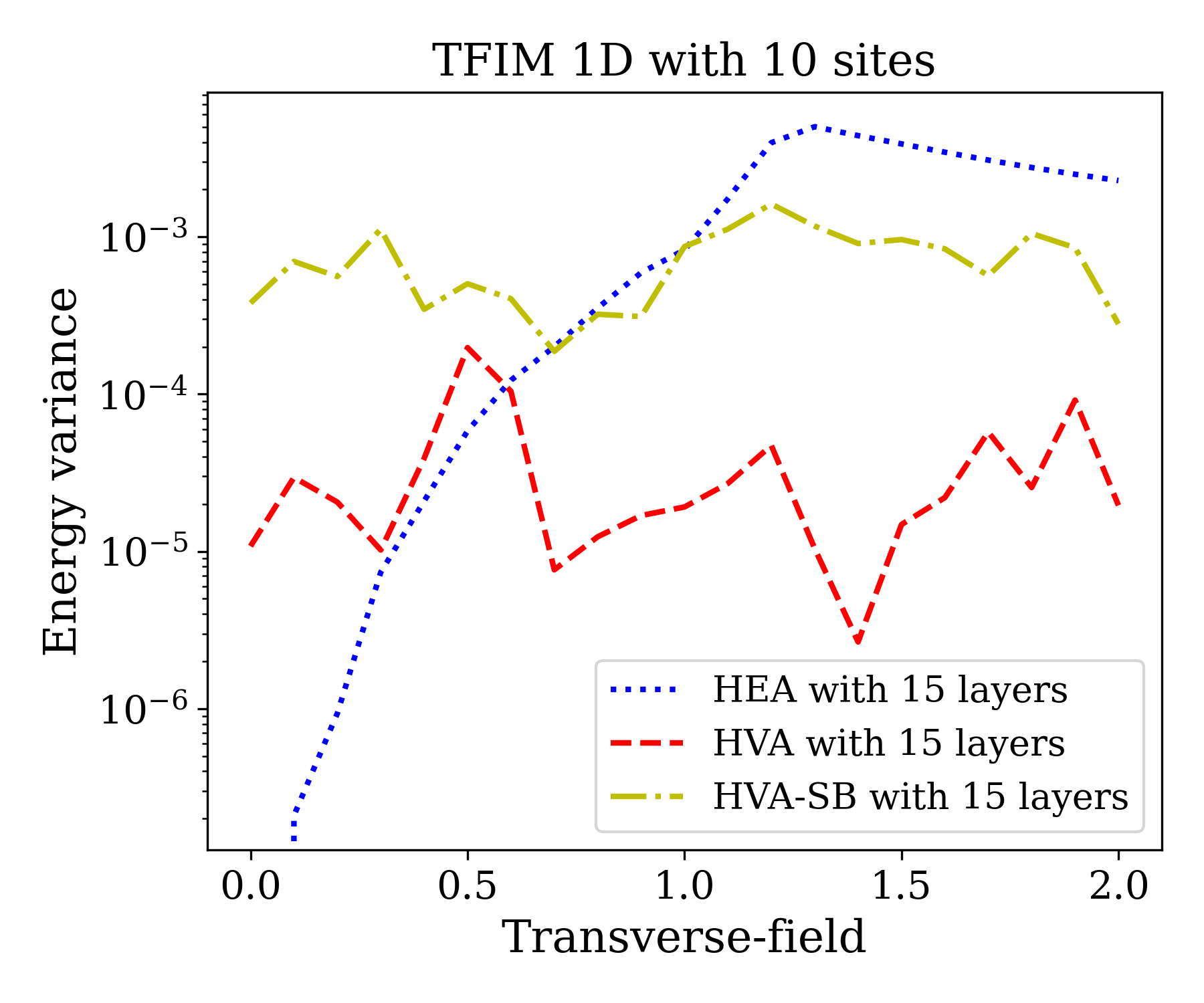}
    \caption{Energy variance vs $h_x$ : TFIM 1-D with 10 sites}
    \label{fig:Var_E_TFIM_1D_10}
\end{figure}

\begin{figure}[htb]
    \centering
    \begin{subfigure}{\linewidth}
        \includegraphics[width = \textwidth]{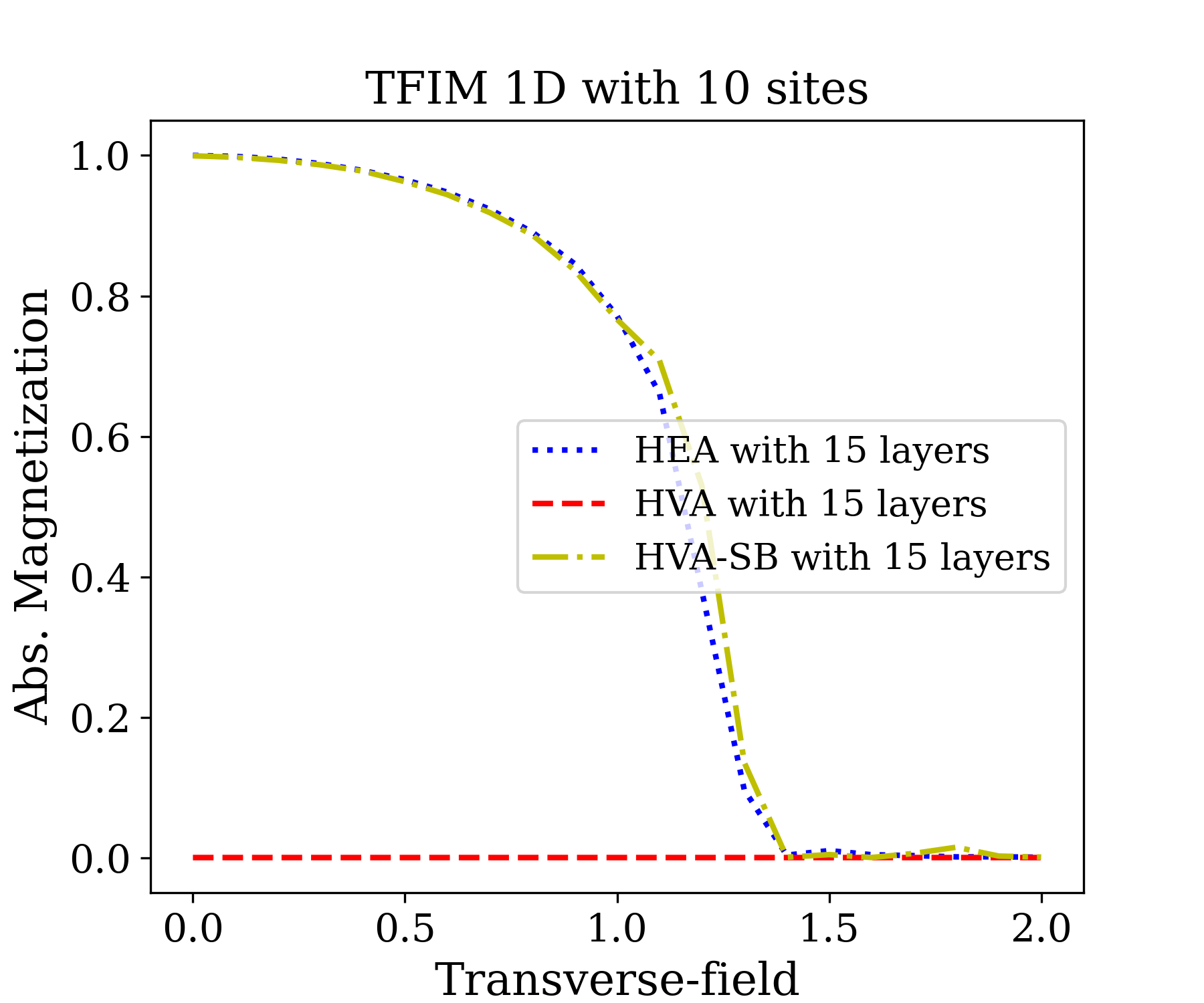}
        \caption{Absolute magnetization}
        \label{fig:1D_Mz_10}
    \end{subfigure}
    \begin{subfigure}{\linewidth}
        \includegraphics[width = \textwidth]{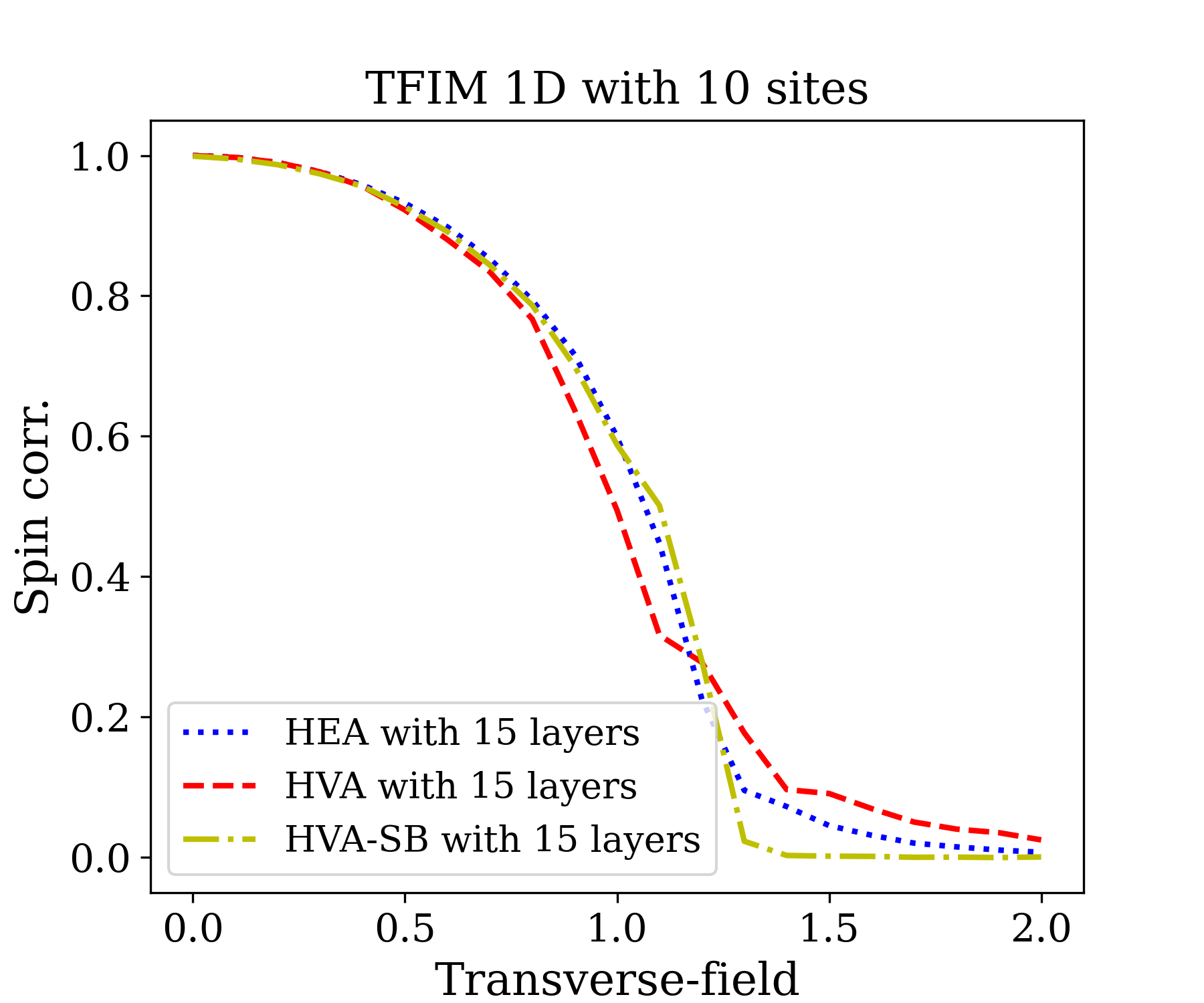}
        \caption{Spin correlation}
        \label{fig:1D_spin_corr_10}
    \end{subfigure}
    \caption{Spin correlation, and Magnetization for 10 site 1D system }
    \label{fig:1D_spin_corr_Mz_10}
\end{figure}

In Fig.~\ref{fig:Var_E_TFIM_1D_10} we show our results for energy variance, as defined in Eq.~\ref{eq:var_E}, across transverse fields for a 10-spin system. The interesting aspect to notice here is that the number of parameters per layer for HEA is 10 (number of sites) times the number of parameters of HVA or HVA-SB. We observe that as one goes from smaller number of layers (parameters) to a higher number of layers, the HEA slightly performs better. However,  in the case of HVA or HVA-SB, there is a sudden jump from a worse result to a better result in comparison to HEA. This jump is due to the restricted Hilbert space of the Hamiltonian variational circuit. As observe, the HVA-SB ans\"atz is less restrictive than the pure HVA. This is due to the inclusion of the symmetry-breaking layer, which allows access to a parity-breaking sector of the Hilbert space.  The symmetry breaking term accesses a less entangled state which only exists in the thermodynamic limit, giving us better energy optimization in a VQE setup.

The variance in energy is represented on a semi-log scale. The figure shows that the variance in the ground-state energy for the HVA circuit is largely insensitive to variations in $h_x$, while for the HEA, the error gradually increases with the widening energy gap between the two lowest eigenstates. This is due to the signature provided by the entanglement graph discussed above, which shows that the HEA provides the superposition state in the highly entangled (low-field) regime. 

We compare these results for the von Neumann entropy shown in Fig.~\ref{fig:1_d_EE_10}. The HEA ans\"atz attempts to reproduce the entanglement entropy of the true ground state; however, as it does so, the energy accuracy deteriorates, reflecting a deviation from the optimal state during optimization. Once the correct entanglement is reached, corresponding to the true ground state, the energy variance becomes nearly constant. In the low-field regime, the optimization is influenced more by the energy gap between nearly degenerate states than by the true ground state itself, particularly when minimizing the energy variance. Although the entanglement entropy is not accurately captured in the HEA and HVA-SB cases, we observe a clear change in the curvature of the entanglement profile near the critical point. This behavior further highlights the importance of choosing an appropriate ans\"atz and an appropriate observable for reliable VQE performance.

The absolute magnetization is shown in Fig.~\ref{fig:1D_Mz_10}. In a finite system, the exact ground state preserves the $\mathbb{Z}_2$ symmetry of the Hamiltonian, resulting in zero magnetization, as discussed in Sec.~\ref{sec:spin_corr}. A non-zero magnetization arises only when the ans\"atz allows superpositions of nearly degenerate states or it effectively breaks the $\mathbb{Z}_2$ symmetry, as observed for the HEA and HVA-SB ans"atze. This leads to an apparent phase transition in the magnetization, even though no true phase transition is expected at finite volume.
In Fig.~\ref{fig:1D_spin_corr_10}, we present the results for the spin correlation. We observe that the spin correlation approaches one-half near the critical point for all ans\"atze. As discussed in Sec.~\ref{sec:spin_corr}, this highlights that spin correlation serves as a more reliable observable for probing criticality in nearly degenerate states within a VQE simulation.

\begin{figure}[htb]
    \centering
    \begin{subfigure}{\linewidth}
        \includegraphics[width = \textwidth]{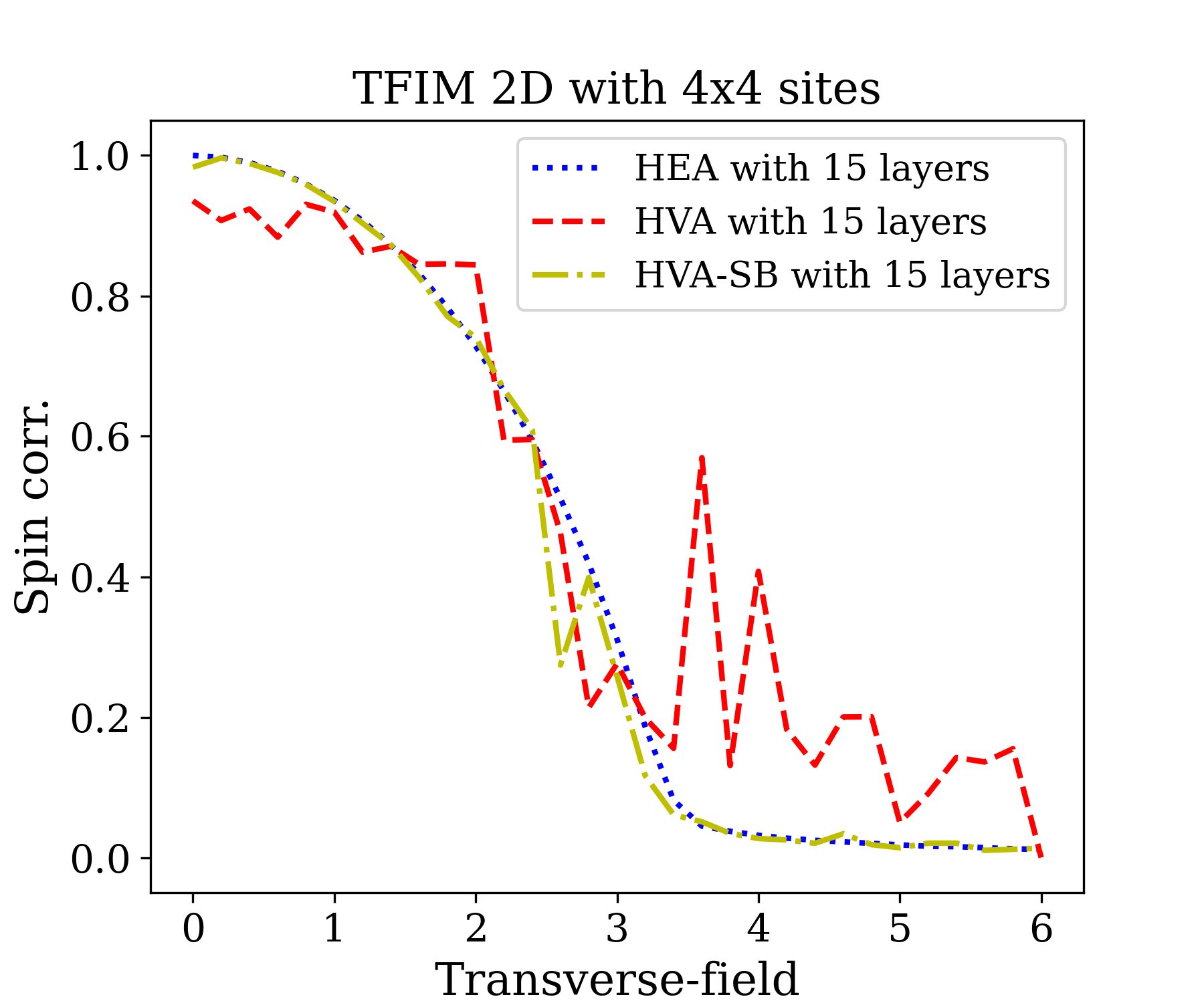}
        \caption{Spin correlation}
    \end{subfigure}
    \begin{subfigure}{\linewidth}
        \includegraphics[width = \textwidth]{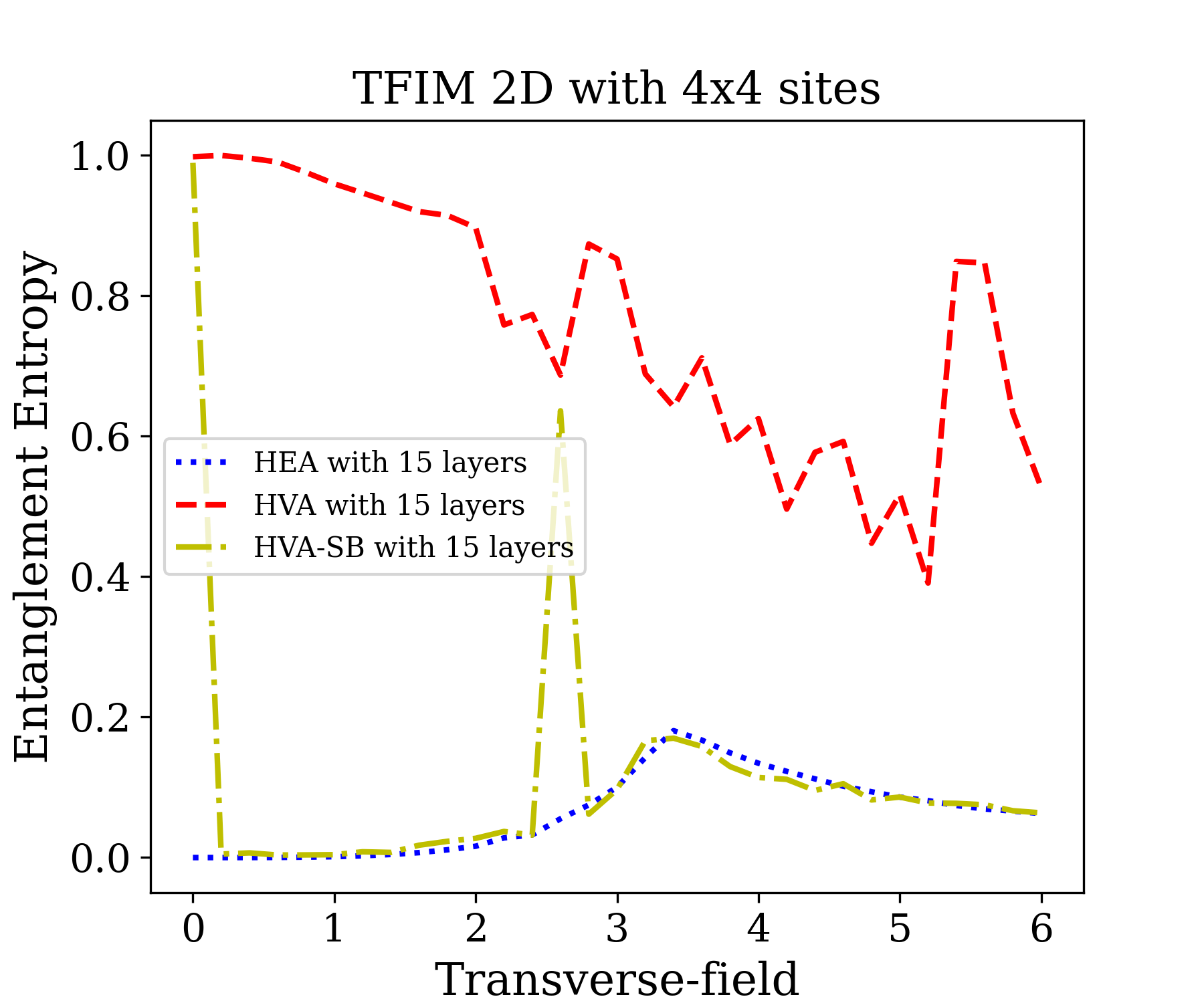}
        \caption{Entanglement entropy}
    \end{subfigure}
    \caption{Spin correlation and Entanglement per site calculation for 2D system of 4x4 dimensions }
    \label{fig:2D_spin_corr_EE}
\end{figure}

We extend our setup to higher dimensions by considering larger systems and applying the same ans\"atz in the two-dimensional case, and results are shown in Fig.~\ref{fig:2D_spin_corr_EE}. However, unlike the one dimensional case,
the instability increases in higher dimensions. This can be attributed in part to the enhanced connectivity in higher dimensions, which makes the optimization landscape more challenging. As shown in Fig.~\ref{fig:2D_spin_corr_EE}, both the spin correlations and energy variance indicate that the HVA performs poorly in low-entanglement regimes, while the HEA underestimates entropy in highly entangled regions. The optimization difficulty is further reflected in the sensitivity to initial parameters, with different initializations leading to different outcomes. However, though a precise quantitative analysis is beyond the scope of this work, both the spin correlations and the von Neumann entanglement entropy exhibit clear signatures of critical behavior near the critical field $h_x \approx 3.3$ for the two-dimensional TFIM. The instability in the computation of the ground state using the HVA ans\"atz increases with system dimensionality, and for that reason, in 3D, it is not possible to benchmark the results. In this case, one needs a much better optimizer with the HVA ans\"atz. The results for the HEA ans\"atz with real amplitudes have already been discussed in Sec.~\ref{sec:results} for 1D, 2D, and 3D systems.

\section{\label{sec:conclusion}Conclusion}
In this work, we have systematically investigated the performance of the Variational Quantum Eigensolver for the transverse-field Ising model across one, two, and three dimensions, with various choices of ans\"atze.  By benchmarking against results obtained by exact diagonalization,  and analyzing observables such as energy, spin correlations, and entanglement entropy, we have assessed the capability of different ans\"atze to capture ground-state properties and critical behavior. Notably, we extend VQE studies to the three-dimensional TFIM for the first time, considering systems as large as 27 spins. 

Our study reveals a trade-off between expressivity and optimization stability in VQE. While hardware-efficient ans\"atze provide smoother optimization landscapes, they fail to accurately capture correlations in strongly entangled regimes. In contrast, Hamiltonian-inspired ans\"atze, including their symmetry-breaking extensions, achieve higher fidelity and better reproduce critical features, albeit with more challenging optimization. The extension to higher dimensions reflects the increasing difficulty of variational optimization in the presence of enhanced connectivity and entanglement.

Overall, our results demonstrate that neither high expressivity nor strict ans\"atz structure alone is sufficient for scalable quantum simulation of many-body systems. Instead, a balanced approach that incorporates physical insight while maintaining optimization tractability is essential. These findings provide guidance for the design of improved variational strategies and underscore the need for adaptive ans\"atz and advanced optimization techniques. Future work along these directions will be crucial for extending VQE to larger system sizes and more complex quantum many body systems.

\section{Acknowledgment}
This work is supported by the Department of Atomic Energy, Government of India, under Project Identification Number RTI-4012.
Computations were carried out on the computing clusters at the Department of Theoretical Physics, TIFR, Mumbai. We are also thankful to Debasish Banerjee for discussions.
We would also like to thank  Ajay Salve and Kapil Ghadiali for computational support.

\bibliographystyle{plainnat}
\bibliography{Quantum_journal}

\appendix

\section{\label{vqe_dis}VQE and its limitations}
Variational quantum eigensolver is a commonly used algorithm in NISQ era taking advantage of quantum scaling while working with noisy quantum hardware. 
It is a hybrid algorithm that uses classical and quantum
processors together. VQE and its variants are generally employed to explore the minimum eigenvalue and corresponding eigenvector of a given quantum system. In general a quantum Hamiltonian is first expressed as,
\begin{equation}
    H = \sum_i c_i h_i,
\end{equation}
where $h_i$ is a tensor product of Pauli matrices and $c_i$ is the coefficient. By optimizing a set of parameters one reaches to the ground state of the system, 
\begin{equation}
    E_0 = \min_{\vec{\theta}} \langle\psi(\vec{\theta)} | H | \psi(\vec{\theta)} \rangle
\end{equation}
In Fig. \ref{fig:vqe_flowchart} we show a representative flowchart of a standard VQE algorithm, comprising both classical and quantum processors.

\begin{figure*}[hbt]
    \centering
    \begin{tikzpicture}[
    font=\sffamily,
    box/.style={draw, thick, rounded corners=6pt, minimum width=6cm, align=center, inner sep=10pt, fill=#1!10},
    arrow/.style={-{Stealth[length=4mm]}, very thick},
    node distance=1cm
    ]
    
    \node[box=gray] (input) {
        \textbf{\Large Input}\\[6pt]
        1. Hamiltonian (Pauli operators)\\
        2. Parametric Quantum Circuit\\
        3. Classical Optimization Method
    };
    
    \node[box=blue, right=of input] (quantum) {
        \textbf{\Large Quantum Processor}\\[6pt]
        Circuit Evaluation\\[4pt]
        $E_n = \langle \psi(\vec{\theta}_n)|H|\psi(\vec{\theta}_n)\rangle$
    };
    
    \node[box=orange, below=of quantum] (classical) {
        \textbf{\Large Classical Processor}\\[6pt]
        Optimizer\\[4pt]
        $\vec{\theta}_{n+1} = F(\{E_n\}, \{\vec{\theta}_n\})$
    };
    
    \node[box=gray, left=of classical] (output) {
        \textbf{\Large Output}\\[6pt]
        1. Optimized Parameters $(\vec{\theta}_0)$\\
        2. Minimum Energy $(E_0)$
    };
    
    \draw[arrow] (input) -- (quantum);
    \draw[arrow] (quantum) -- node[right=4pt]{ $E_n, \vec{\theta}_n$ } (classical);
    \draw[arrow] (classical) -- (output);
    
    \draw[arrow]
        (classical.east) to[out=0, in=0, looseness=.8]
        node[left=4pt]{ $\vec{\theta}_{n+1}$ }
        (quantum.east);
    
    \end{tikzpicture}
    
    \caption{VQE flowchart}
    \label{fig:vqe_flowchart}
\end{figure*}
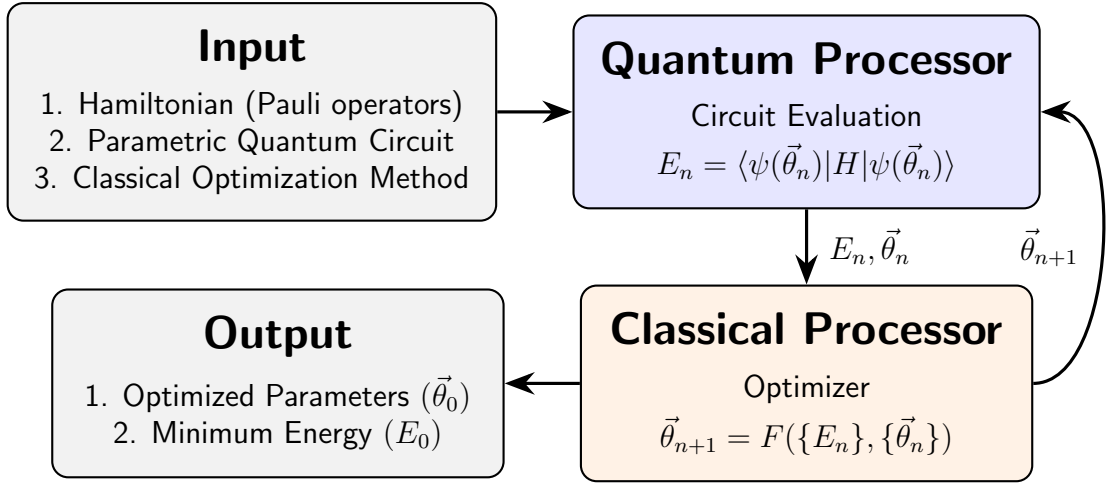

\subsection{Expressivity of a VQE circuit}

Though VQE is a general algorithm, one needs to be careful about the expressivity of a given VQE circuit. Below we explain that with a simple example. Imagine we want to study a 2-qubit Hamiltonian with only one term,
\begin{equation}
    H = \sigma^x_0\sigma^z_1
\end{equation}
One can use various ans\"atz circuits and we show two of them in Fig.\ref{fig:Circuit1} (ans\"atz-1) and \ref{fig:Circuit2} (ans\"atz-2). Both ans\"atze use only one parameter $\theta$. By defining the energy function,
\begin{equation}
    E(\theta) = \bra{\psi(\theta)}  H \ket{\psi(\theta)},
\end{equation}
for ans\"atz-1, one arrives
\begin{equation}
    \scriptsize
    \begin{split}
        E_1(\theta) & = \left( \cos{\frac{\theta}{2}} \bra{00} + \sin{\frac{\theta}{2}} \bra{11} \right) \sigma^x_0\sigma^z_1 \left( \cos{\frac{\theta}{2}} \ket{00} + \sin{\frac{\theta}{2}} \ket{11} \right) \\
        & = \left( \cos{\frac{\theta}{2}} \bra{00} + \sin{\frac{\theta}{2}} \bra{11} \right) \left( \cos{\frac{\theta}{2}} \ket{10} - \sin{\frac{\theta}{2}} \ket{01} \right) \\
        & = 0
    \end{split}
\end{equation}

\begin{figure}[hbt]
    \centering
    \begin{subfigure}{0.4\linewidth}
        \includegraphics[width = \linewidth]{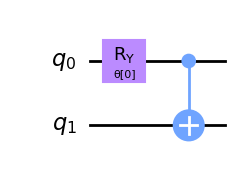}
        \caption{}
        \label{fig:Circuit1}
    \end{subfigure}
    \begin{subfigure}{0.3\linewidth}
        \includegraphics[width = \linewidth]{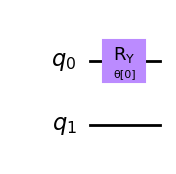}
        \caption{}
        \label{fig:Circuit2}
    \end{subfigure}
    \caption{VQE circuits}
\end{figure}
As we can see that the above energy function does not depend on any parameter. In this case, ans\"atz-1 cannot represent the ground state of the given Hamiltonian. However, for ans\"atz-2 the corresponding energy function is given by,
\begin{equation}
\scriptsize
    \begin{split}
        E_2(\theta) & = \left( \cos{\frac{\theta}{2}} \bra{00} + \sin{\frac{\theta}{2}} \bra{10} \right) \sigma^x_0\sigma^z_1 \left( \cos{\frac{\theta}{2}} \ket{00} + \sin{\frac{\theta}{2}} \ket{10} \right) \\
        & = \left( \cos{\frac{\theta}{2}} \bra{00} + \sin{\frac{\theta}{2}} \bra{10} \right) \left( \cos{\frac{\theta}{2}} \ket{10} + \sin{\frac{\theta}{2}} \ket{00} \right) \\
        & = 2 \sin{\frac{\theta}{2}} \cos{\frac{\theta}{2}} \\
        & = \sin{\theta}
    \end{split}
\end{equation}
In this case, the energy function depends on the parameter $\theta$ and also can provide the correct minimum eigenvalues: -1 for $\theta = -\pi/2$.

\section{\label{vqe_resource} Resource Estimation}
An ans\"atz can be classified by many properties, namely the number of parameters, CNOT gates, and circuit depth. The number of layers is denoted by $N_L$, and the number of qubits by $N_Q$. Since the HVA and HVA-SB ans\"atze depend on the Hamiltonian, the resource estimation also depends on dimensions $d$. 
The higher expressivity of the HEA comes from the fact that it has more parameters, as well as from its design, which is more efficient in terms of local rotations and entanglement. Since every qubit is entangled with every other qubit, the circuit depth does not depend only on the number of layers but also on the number of qubits. Also, every rotation gate has independent parameters, making the number of parameters depend on the number of qubits as well.

The HVA and HVA-SB ans\"atz collect all commuting terms under a single parameter (i.e., depend on a single parameter), and the number of parameters depends only on the number of layers. The CNOT gates depend only on the local interactions in the lattice, and hence the circuit depth depends only on the number of layers, $N_L$, and the dimension $d$, while the total number of CNOT gates depends on $N_Q$. In Table~\ref{tab:ansatz_properties}, we provide a comparison between the HEA, HVA, and HVA-SB ans\"atze for TFIM across dimensions. 

We compare the performance of different circuits. Here, we present statevector and finite-shot (statistical noise) simulations of the HEA and HVA circuits in Tables \ref{tab:statevector_sim} and \ref{tab:statistical_sim} for the 1D TFIM with 10 sites (qubits). As can be seen, the HVA circuit not only requires fewer parameters but also achieves a better ground-state energy with fewer circuit evaluations, resulting in faster convergence.
\begin{table*}[htb]
    \centering
    \begin{tabular}{|l|c|c|c|}
        \hline
         ans\"atz Circuit & parameters & CNOT gates & circuit depth\\
         \hline 
         HEA & $2 (N_L+1) N_Q$ & $ 2 N_L N_Q (N_Q-1)$ & $N_L (N_Q+1) + 2$ \\
         HVA & $2 N_L$ & $ 2d N_L N_Q $ & $ (2d+1) N_L + 1$ \\
         HVA-SB & $3 N_L$ & $ 2d N_L N_Q $ & $ 2(d+1) N_L + 1$ \\
         \hline 
    \end{tabular}
    \caption{Resource estimation corresponding to HEA, HVA and HVA-SB ans\"atze for 1D, 2D and 3D TFIM}
    \label{tab:ansatz_properties}
\end{table*}

\begin{table*}[hbt]
    \centering
    \begin{tabular}{|c|c|c|c|c|c|c|}
        \hline
        Ansatz & layers & parameters & optimizer & total evals & energy & time (min) \\
        \hline
        HEA-full & 10 & 220 & COBYLA & 100000 &  -12.5952 & 50.41 \\
        HEA-linear & 10 & 220 & COBYLA & 100000 &  -12.4714 & 46.30\\
        HVA & 10 & 20 & COBYLA & 4832 & -12.78488 & 1.05\\
        \hline
    \end{tabular}
    \caption{Statevector simulation}
    \label{tab:statevector_sim}
\end{table*}

\begin{table*}[hbt]
    \centering
    \begin{tabular}{|c|c|c|c|c|c|c|}
        \hline
        Ansatz & layers & parameters & optimizer & total evals & energy & time (min) \\
        \hline
        HEA-full & 10 & 220 & COBYLA & 100000 &  -12.65204 & 29.45\\
        HEA-linear & 10 & 220 & COBYLA & 100000 &  -12.65727 & 24.75\\
        HVA & 10 & 20 & COBYLA & 4832 & -12.78486 & 0.7 \\
        \hline
    \end{tabular}
    \caption{Statistical simulation (shots = 1024)}
    \label{tab:statistical_sim}
\end{table*}

\section{\label{ED_DMRG_VQE_compare} Comparison of ED, DMRG and VQE methods}
The ground state of a Hamiltonian can be computed numerically using several approaches, with two widely used classical methods being exact diagonalization (ED) and tensor network (TN) techniques. While ED is limited to small system sizes due to the exponential growth of Hilbert space and associated memory constraints, TN methods particularly in one dimension enable the study of significantly larger finite systems. Here, we employ the density matrix renormalization group (DMRG) within the tensor network framework \cite{Schollw_ck_2011, Catarina_2023}. 
As a test case, for the one-dimensional transverse-field Ising model (TFIM) with 10 spins, we have presented a comparison of results obtained using exact diagonalization (ED), DMRG (MPS), and the variational quantum eigensolver (VQE) in Fig. 
\ref{fig:1_d_energy_10}. In Table \ref{tab:1D_ED_DMRG_VQE}, we show the numerical values obtained for energy and entanglement entropy at different magnetic field ($h_x$) using various numerical methods. Notably, even short-range entangling gates in VQE are capable of capturing long-range entanglement and reproducing the exact ground state with an appropriate ans\"atz. We find that the Hamiltonian Variational ans\"atz (HVA) shows close agreement with ED results. At zero transverse field, the system exhibits two degenerate ground states, and the true ground state can be expressed as a linear combination of these. While ED typically yields the two simplest computational basis states, the HVA ans\"atz within the VQE framework preserves the underlying $\mathbb{Z}_2$ symmetry.

\begin{table*}[hbt]
    \centering
    \begin{tabular}{|r|r|r|r|r|r|r|}
        \hline
        & \multicolumn{3}{|c|}{Energy} &  \multicolumn{3}{|c|}{Entropy}  \\
        \hline
        ~$ h_x $~ & ~~~~~ED~~~~~ & ~~~DMRG~~~ & ~~~~VQE~~~~ & ~~~~~ED~~~~~ & ~~~DMRG~~~ & ~~~~VQE~~~~ \\
        \hline
        0.0 & -10.00000 & -10.00000 & -9.99981 & -0.00000 & -0.00000 &  1.00000 \\ 
        0.1 & -10.02502 & -10.02502 & -10.02430 &  0.99819 &  0.00002 &  0.99824 \\ 
        0.2 & -10.10025 & -10.10025 & -10.09960 &  0.99270 &  0.00023 &  0.99295 \\ 
        0.3 & -10.22630 & -10.22630 & -10.22600 &  0.98332 &  0.00110 &  0.98330 \\ 
        0.4 & -10.40419 & -10.40417 & -10.40320 &  0.96968 &  0.00314 &  0.96899 \\ 
        0.5 & -10.63560 & -10.63545 & -10.62730 &  0.95107 &  0.01270 &  0.95025 \\ 
        0.6 & -10.92331 & -10.92302 & -10.91780 &  0.92613 &  0.55779 &  0.92322 \\ 
        0.7 & -11.27223 & -11.27186 & -11.27200 &  0.89202 &  0.78889 &  0.89007 \\ 
        0.8 & -11.69094 & -11.69093 & -11.68990 &  0.84319 &  0.84283 &  0.84778 \\ 
        0.9 & -12.19187 & -12.19187 & -12.19150 &  0.77259 &  0.77259 &  0.77535 \\ 
        1.0 & -12.78491 & -12.78491 & -12.78440 &  0.68090 &  0.68090 &  0.68707 \\ 
        1.1 & -13.46610 & -13.46610 & -13.46520 &  0.58433 &  0.58433 &  0.57603 \\ 
        1.2 & -14.21775 & -14.21775 & -14.21280 &  0.49995 &  0.49995 &  0.52155 \\ 
        1.3 & -15.02016 & -15.02016 & -15.01890 &  0.43249 &  0.43249 &  0.44426 \\ 
        1.4 & -15.85851 & -15.85851 & -15.85850 &  0.37934 &  0.37934 &  0.37580 \\ 
        1.5 & -16.72302 & -16.72302 & -16.72170 &  0.33682 &  0.33682 &  0.34551 \\ 
        1.6 & -17.60727 & -17.60727 & -17.60500 &  0.30204 &  0.30204 &  0.31280 \\ 
        1.7 & -18.50690 & -18.50690 & -18.50410 &  0.27304 &  0.27304 &  0.28152 \\ 
        1.8 & -19.41884 & -19.41884 & -19.41680 &  0.24846 &  0.24846 &  0.25413 \\ 
        1.9 & -20.34085 & -20.34085 & -20.33550 &  0.22735 &  0.22735 &  0.23808 \\ 
 2.0 & -21.27121 & -21.27121 & -21.26960 &  0.20904 &  0.20904 &  0.21524 \\ 
        \hline
    \end{tabular}
    \caption{A comparison between ED, DMRG and VQE methods for energy and entanglement entropy}
    \label{tab:1D_ED_DMRG_VQE}
\end{table*}

\section{Discussion on Hardware implementation}
Hardware connectivity plays a crucial role in determining the efficiency of quantum simulations, particularly for lattice Hamiltonians with local interactions such as the transverse-field Ising model (TFIM). For one-dimensional TFIM, nearest-neighbor interactions naturally map onto linear qubit arrays, making the underlying hardware topology relatively unimportant. In contrast, two- and three-dimensional TFIM require hardware connectivity that closely matches the lattice geometry in order to minimize the routing overhead introduced by SWAP operations.

Current quantum hardware already provides several architectures that are well suited for two-dimensional lattice simulations. Superconducting processors from IBM employ either the heavy-hex lattice (e.g., the Heron family) or the square-lattice architecture (Nighthawk) \cite{IBMProcessorTypes}. While the heavy-hex topology provides each qubit with two or three nearest neighbors, Nighthawk increases this to four nearest neighbors, making it particularly attractive for square-lattice Hamiltonians such as the 2D TFIM. Consequently, local interactions can be embedded with at most two to three degrees of separation, substantially reducing the compilation overhead compared with less compatible hardware topologies.

Trapped-ion platforms offer an alternative approach. Because any pair of ions can, in principle, be entangled through collective motional modes, they provide flexible connectivity that is also well suited for two-dimensional local Hamiltonians. Furthermore, scalable trapped-ion architectures based on ion shuttling have been proposed, where ions are physically transported between interaction regions instead of relying exclusively on logical SWAP gates. Although ion transport introduces its own time overhead, this cost is generally less detrimental than the rapidly increasing circuit depth associated with repeated SWAP operations on sparsely connected architectures, making shuttling an attractive strategy for large-scale quantum simulations.

While the three-dimensional lattice case would be less challenging for the trapped ion with classical shuttling, simulating it on a superconducting qubit would present a greater challenge. A cubic lattice requires each bulk site to interact with six nearest neighbors, exceeding the connectivity available on current superconducting processors. IBM has proposed future processor architectures incorporating up to six-way connectivity, which would enable a much more direct embedding of three-dimensional lattice Hamiltonians.

These developments are particularly relevant to the TFIM benchmarks presented in this work. While our simulations assume ideal logical qubits without hardware constraints, the required interaction graphs are already well aligned with existing two-dimensional quantum processors and are expected to become increasingly compatible with future hardware. Together with advances in qubit connectivity and trapped-ion shuttling technologies, these developments indicate that large-scale quantum simulations of lattice spin models, including higher-dimensional TFIM, are becoming progressively more practical on near-term quantum hardware.
\end{document}